\documentclass[aps,prd,nofootinbib]{revtex4}
\usepackage{graphicx}
\usepackage{hyperref}
\usepackage{slashed}
\usepackage{color}
\usepackage{subfigure}

\begin{document}

\title {Production double heavy quarkonium at Z mass pole}

\author{Qi-Li Liao}
\email{xiaosueer@163.com}
\affiliation{Chongqing College of Mobile Telecommunications, Chongqing 401520, China}

\date{\today}

\begin{abstract}
The exclusive production of double excited quarkonium is comprehensive studied, i.e., the production of double excitedcharmonium, double excited bottomonium, and double excited $Bc$-mesons via $e^+e^-\to \gamma^*/Z^0 \to|(Q\bar{Q'})[n]\rangle +|(Q'\bar{Q})[n']\rangle$ ($Q/Q'=c$- or $b$-quarks) at a future $Z$ factory under the nonrelativistic quantum chromodynamics framework, where the $[n]$~/$[n']$ represents the color-singlet heavy quarkonium states $[^1S_0]\rangle, ~[^3S_1]\rangle, ~[^1P_1]\rangle$, and $[^3P_J]\rangle$ ($J=0,1,2$).
The ``improved trace technology" is adopted for calculating the complicated $P$-wave channels for derive the analytic expressions at the amplitude level.
According to our study, the production rates of double heavy quarkonium are considerable at the future $Z$ factory.
We obtain the cross sections for the production of double  excited charmonium for $\sigma{(|(c\bar{c})[n]\rangle+|(c\bar{c})[n']\rangle)_{total}}=1.167^{+0.113}_{-0.164}\times 10^{-2}~fb$, the cross sections of double  excited bottomonium for $\sigma{(|(b\bar{b})[n]\rangle+|(b\bar{b})[n']\rangle)_{total}}=0.1132^{+0.0139}_{-0.0130}~fb$, and the cross sections of double excited $Bc$-mesons for $\sigma{(|(c\bar{b})[n]\rangle+|(b\bar{c})[n']\rangle)_{total}}=3.692^{+0.124}_{-0.097}~fb$.
The main uncertainties come from the mass of the heavy quarkonium and the radial wave functions at the origin and their derivatives at the origin under different potential models.
The numerical results show that such the super $Z$ factory should be a good platform to study the properties of the double excited charmonium, the double excited bottomonium , especially the double excited $Bc$-mesons.
\end{abstract}

\maketitle

\section{Introduction}
In comparison to the hadronic colliders like Large Hadron Collider (LHC), an electron-positron collider has many advantages, as it provides a precise collision energy and a cleaner hadronic background. And polarization of incoming positron and electron beams can be well controlled.
A super $Z$ factory running at the energy of the $Z^0$-boson mass with high luminosity ${\cal L}\approx 10^{34\sim36}cm^{-2}s^{-1}$ has been proposed \cite{jz}, which is similar to the GigaZ mode at the Circular Electron-Positron Collider (CEPC) \cite{CEPCStudyGroup:2018ghi} and an Electron-Positron Linear Collider \cite{ECFADESYLCPhysicsWorkingGroup:2001igx}.
Due to the high yields of $Z^0$ bosons up to $7\times 10^{11}$ at the CEPC \cite{CEPCStudyGroup:2018ghi}, it can be used for studying the production of double heavy quarkonium through $Z^0$ decays.

As the heavy quarkonium is a multiscale problem for probing quantum chromodynamics (QCD) theory at all energy regions. That provides an ideal platform to investigate the properties of bound states.
The production of double heavy quarkonium has been studied extensively both at the LHC and $B$ factories.
The hadronic production of double $J/\psi$ has always been a hot topic. Since $J/\psi$ can be very easy to be detected by its leptonic decays, and this channel can be used to explore the distribution of gluons in a proton at the LHC \cite{Schafer:2019ynn,Scarpa:2019fol,He:2019qqr,Qiao:2002rh,Qiao:2009kg,Lansberg:2013qka,Lansberg:2020rft,Lu:2021gxf}.
In addition, the production of double heavy quarkonium through photon-photon interaction \cite{Chen:2020dtu,Yang:2020xkl}, the production of double heavy quarkonium through diffractive interactions \cite{BrennerMariotto:2018eef}, the photoproduction of double $J/\psi$ \cite{Xue-An:2018wat}, and the hadronic production of double $B_c$ mesons \cite{Li:2009ug} have also been studied.
The production of double heavy quarkonium through positron-electron annihilation has be explored at $B$ factories.
In particular, the production of $J/\psi + \eta_c$ at $B$ factories once challenged the nonrelativistic QCD (NRQCD)  \cite{nrqcd1,nrqcd2}. Considering the fact of the nonrelativistic nature of heavy quark and antiquark inside the quarkonium, the NRQCD could be a powerful tool to study the production and decay mechanism of heavy quarkonium.
At the leading order under the NRQCD formulation, its total cross section is about $2\sim 6$ fb \cite{Braaten:2002fi,Liu:2002wq,Hagiwara:2003cw}.
However, the measurements at $B$ factories by the Belle and BaBar collaborations show that the cross section is about 20 fb \cite{Belle:2002tfa,BaBar:2005nic}.
This large discrepancy between experiment and theory can be reduced by the QCD next-to-leading order corrections \cite{Zhang:2005cha,Gong:2007db} and the relativistic corrections in the NRQCD \cite{Braaten:2002fi,He:2007te,Bodwin:2007ga}.
The production total cross section of $J/\psi + \eta_c$ by the QCD next-to-next-to-leading corrections final gives consistent estimate with the BaBar measurement  \cite{Feng:2019zmt}.

As the analytical expressions for the usual squared amplitudes in short-distance coefficients becomes complicated and lengthy for massive particles in final states,  especially for processes involving the $P$-wave Fock states. To solve the problem, the ``improved trace technology" is suggested and developed \cite{cjx,lxz,Yang:2011ps,wbc1,Liao:2012rh,lx,Liao:2015vqa,lx1,lx2,qjpg}, which is based on the helicity amplitudes method and deals with the trace calculation directly at the amplitude level.
In this way, the amplitudes could be expressed with the linear combinations of independent Lorentz structures.
In this paper, the ``improved trace technology" is also adopted to derive the analytical expression for all processes.

In previous works \cite{gxz1,gxz2}, the production of ground states ($1S$ and $1P$-wave) charmonium in $e^+e^-\to \gamma^*/Z^0 \to|(c\bar{c})\rangle +|(c\bar{c})\rangle$ at the super $Z$ factory are studied at the leading order and next-to-leading order in strong coupling constant $\alpha_s$ under the NRQCD framework. The associated $S$-wave charmonium-bottomonium production at LO at $Z^0$ pole are explored \cite{Belov:2021ftc}. The QCD NLO corrections to the production of $J/\psi+J/\psi$ and $J/\psi+\eta_c$ are important near the $Z^0$ mass pole \cite{Berezhnoy:2021tqb}.
The QCD NLO corrections to the paired $S$-wave $B_c^{(*)}$ production are small at $Z^0$ mass pole \cite{Berezhnoy:2016etd}.
In the present paper, we shall concentrate our attention on the production of both ground and excited Fock states of double charmonium, double bottomonium, and double $B_c$ mesons in $e^+e^-\to \gamma^*/Z^0 \to|(Q'\bar{Q})[n]\rangle +|(Q\bar{Q'})[n']\rangle~$ ($Q/Q'=c$-,$b$-quark) at the future super $Z$ factory, where $[n]$~/$[n']$ represents the color-singlet $[^1S_0]\rangle, ~[^3S_1]\rangle, ~[^1P_1]\rangle$, and $[^3P_J]\rangle$ ($J=0,1,2$) heavy quarkonium states.
Then we shall discuss differential angle distribution, the transverse momentum distribution, and the uncertainties caused by the masses of constituent heavy quarks and the nonperturbative long-distance matrix elements.
The phenomenological analysis would be a helpful support for the experimental exploration on the production of double charmonium, double bottomonium, and double $B_c$ mesons at the future super $Z$ factory or GigaZ mode at CEPC.

The rest of the present paper is organized as follows.
In Sec. II, we introduce the calculation formalism for the processes of $e^+e^-\to \gamma^*/Z^0 \to|(Q\bar{Q'})[n]\rangle +|(Q'\bar{Q})[n']\rangle$ under the NRQCD factorization framework.
In Sec. III,  we evaluate the cross sections. The differential distributions of the cross sections and the uncertainties from various sources are studied in Sec. \ref{production} and \ref{uncertainty}, respectively.
The final Sec. IV is reserved for a summary.

\section{Formulations and Calculation Techniques}

In the NRQCD framework, it divides the calculation into the long-distance matrix elements and the short-distance coefficients. The long-distance matrix elements describe the hadronization of Fock states with $J^{PC}$ quantum numbers into heavy quaronium and are nonperturbative parameters. The short-distance coefficients describe the hard scattering of partons and can be calculated perturbatively via Feynman diagrams.

The cross sections for the production of the double excited quarkonium in $e^+e^-\to \gamma^*/Z^0 \to |(Q'\bar{Q})[n]\rangle +|(Q\bar{Q'})[n']\rangle$ can be calculated under the NRQCD factorization framework \cite{gxz1}.
The differential cross sections can be factored into  the long-distance matrix elements and the short-distance coefficients,
\begin{equation} \label{dsigma}
d\sigma=\sum  d\hat\sigma(|(Q'\bar{Q})[1]\rangle+ |(Q\bar{Q'})[1]\rangle) {\langle{\cal O}^H(n) \rangle}{\langle{\cal O}^H(n') \rangle}.
\end{equation}
Here the long-distanc non-perturbative NRQCD matrix elements $\langle{\cal O}^{H}(n)\rangle$~/$\langle{\cal O}^{H}(n')\rangle$ describes the hadronization of a Fock state $(Q\bar{Q'})$~/$(Q'\bar{Q})$, $(Q\bar{Q'})$~/$(Q'\bar{Q})$ into the heavy quarkonium $|(Q\bar{Q'})[n]\rangle$~/$|(Q'\bar{Q})[n']\rangle$. $\hat\sigma(|(Q'\bar{Q})[n]\rangle+ |(Q\bar{Q'})[n']\rangle)$ describes the short-distance production of a $(Q\bar{Q'})~/(Q'\bar{Q})$ pair ($Q/Q'=c$- or $b$-quarks) in the color, spin, and angular momentum states $[n]$~/$[n']$.
Were the $[n]$~/$[n']$ represents the color-singlet $[^1S_0]\rangle, ~[^3S_1]\rangle, ~[^1P_1]\rangle$, and $[^3P_J]\rangle$ ($J=0,1,2$) heavy quarkonium.

The color-singlet nonperturbative matrix element $\langle{\cal O}^H(n) \rangle$~/$\langle{\cal O}^H(n') \rangle$ can be related either to the Schr${\rm \ddot{o}}$dinger wave function $\psi_{(Q\bar{Q'})}(0)$ at the origin for the $S$-wave quarkonium or the first derivative of the wave function $\psi^\prime_{(Q\bar{Q'})}(0)$ at the origin for the $P$-wave quarkonium:

\begin{eqnarray}
\langle{\cal O}^H(1S) \rangle &\simeq& |\Psi_{\mid(Q\bar{Q'})[1S]\rangle}(0)|^2,\nonumber\\
\langle{\cal O}^H(1P) \rangle &\simeq& |\Psi^\prime_{\mid(Q\bar{Q'})[1P]\rangle}(0)|^2.
\end{eqnarray}
Due to the fact that the spin-splitting effects are small, the same values of wave function for both the spin-singlet and spin-triplet Fock states are adopted in our calculation.
Further, the Schr\"{o}dinger wave function at the origin $\Psi_{|Q\bar{Q'})[1S]\rangle}(0)$ and its first derivative at the origin $\Psi^{'}_{|(Q\bar{Q'})[1P]\rangle}(0)$ are related to the radial wave function at the origin $R_{|(Q\bar{Q'})[1S]\rangle}(0)$ and its first derivative at the origin $R^{'}_{|(Q\bar{Q'})[P]\rangle}(0)$, respectively \cite{nrqcd1},
\begin{eqnarray}
\Psi_{|(Q\bar{Q'})[1S]\rangle}(0)&=&\sqrt{{1}/{4\pi}}R_{|(Q\bar{Q'})[1S]\rangle}(0),\nonumber\\
\Psi'_{|(Q\bar{Q'})[1P]\rangle}(0)&=&\sqrt{{3}/{4\pi}}R'_{|(Q\bar{Q'})[1P]\rangle}(0).
\end{eqnarray}

\begin{figure}
\includegraphics[width=0.4\textwidth]{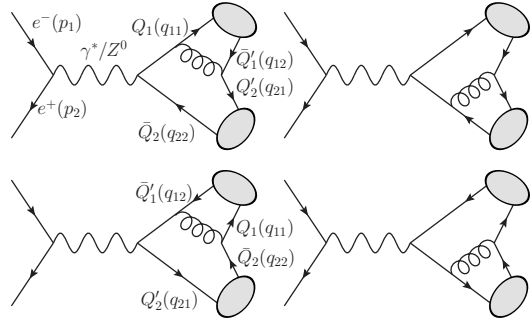}
\caption{Feynman diagrams for processes of $e^-(p_1) e^+(p_2) \to |(Q\bar{Q}')[n]\rangle (q_1) + |(Q'\bar{Q})[n']\rangle (q_2)$, where $|(Q\bar{Q}')[n]\rangle~/
|(Q'\bar{Q})[n']\rangle$ stands for $|(c\bar{c})[n]\rangle$~/ $|(c\bar{c})[n']\rangle$, $|(b\bar{b})[n]\rangle$~/$|(b\bar{b})[n']\rangle$, and $|(c\bar{b})[n]\rangle~/|(b\bar{c})[n']\rangle$ quarkonium.
Here $[n]$~/$[n']$ is short for the color-singlet $[^1S_0]\rangle, ~[^3S_1]\rangle, ~[^1P_1]\rangle$, and $[^3P_J]\rangle$ ($J=0,1,2$) heavy quarkonium.} \label{feyn1}
\end{figure}

The short-distance differential cross section $d\hat\sigma$ are perturbatively calculable, and the four Feynman diagrams of the processes of $e^-(p_1)e^+(p_2)\to \gamma^*/Z^0 \to|(Q\bar{Q'})[n]\rangle(q_1) +|(Q'\bar{Q})[n']\rangle(q_2)$ are displayed in Fig.~\ref{feyn1}.
The perturbative differential cross section can be expressed as
\begin{eqnarray} \label{sd-sigma}
d\hat\sigma=\frac{1}{4\sqrt{(p_1\cdot p_2)^2-m^4_{e}}} \overline{\sum}  |{\cal M}(n)|^{2} d\Phi_2,
\end{eqnarray}
where $\overline{\sum}$ stands for the average over the spin of the initial particles, and sum over the color and spin of the final particles when manipulating the squared amplitudes $|M(n)|^{2}$.
In the $e^-e^+$ center-of-momentum (CM) frame, the two-body phase space can be simplified as
\begin{eqnarray}
    d{\Phi_2} &=& (2\pi)^4 \delta^{4}\left(p_1+p_2 - \sum_{f=1}^2 q_{f}\right)\prod_{f=1}^2 \frac{d^3{\vec{q}_f}}{(2\pi)^3 2q_f^0} \nonumber \\
    &=& \frac{{\mid\vec{q}_1}\mid}{8\pi\sqrt{s}}d(cos\theta).
    \label{dPhi-2}
\end{eqnarray}
Where the parameter $s=(p_1+p_2)^2$ stands for the squared CM energy, and $\theta$ is the angle between the momentum $\vec{p_1}$ of electron and the momentum $\vec{q_1}$ of heavy quarkonium. The magnitude of the 3-dimension quarkonium momentum is ${|\vec{q}_1|}=\sqrt{\lambda[s, M^2_{Q\bar{Q'}}, M^2_{Q'\bar{Q}}]} /2\sqrt{s}$, where $\lambda[a, b, c]= (a - b - c)^2 - 4 c b$, and $M_{Q\bar{Q'}}/M_{Q'\bar{Q}}$ is the mass of heavy quarkonium.
The hard scattering amplitude ${\cal M}(n)$ in Eq. (\ref{sd-sigma}) from the Feynman diagrams in Fig.~\ref{feyn1} can be formulated as

\begin{eqnarray} \label{amplitude}
i{\cal M}(n)=  \sum_{k=1}^4 \bar{v}_{s'} (p_2) {\cal L}^\mu u_s (p_1)  {\cal D}_{\mu\nu} {\cal A}^\nu_k,
\end{eqnarray}
Where the index $k$ represents the number of the Feynman diagrams, and $s$ and $s'$ are the spins of the initial particles.
The vertice $\cal{L}^\mu$ and the propagator $\cal{D_{\mu\nu}}$ for the virtual photon and $Z^0$ propagated processes have different forms,
\begin{eqnarray}
    \cal{L}^\mu &= \left\{
    \begin{array}{l}
        -i e \gamma^\mu \\
        \frac{-i g} {4 cos\theta_W} \gamma^\mu (1-4 sin^2 \theta_W-\gamma^5),
    \end{array}
    \right. \\
    \cal{D_{\mu\nu}} &= \left\{
    \begin{array}{l}
        \frac{-i g_{\mu\nu}}{k^2} \\
        \frac{i}{k^2-m^2_Z+i m_Z \Gamma_Z}(-g_{\mu\nu}+ \frac{k_\mu k_\nu }{k^2}).
    \end{array}
    \right.
\end{eqnarray}
Here the upper and lower expressions after the big left bracket are for virtual photon and $Z^0$ propagated processes, respectively. $e$ is the unit of the electric charge, $g$ is the weak interaction coupling constant, $\theta_W$ represents the Weinberg angle, and $m_Z$ and $\Gamma_Z$ are the mass and the total decay width of $Z^0$ boson, respectively.

The explicit expressions of the Dirac $\gamma$ matrix chains ${\cal A}^\nu_k$ in Eq. (\ref{amplitude}) for the double $S$-wave spin-singlet $^1S_0$ and spin-triplet $^3S_1$ states, and the double $S$-waves spin-triplet $^3S_1$ states from the Feynman diagrams in Fig.~\ref{feyn1} can be formulated as
\begin{widetext}
\begin{eqnarray}
{\cal A}^{\nu (S=0, L=0)}_1 &=& i Tr \left[\Pi^0_{q_1}(0) \gamma^\sigma  \frac{(\slashed{q}_1+ \slashed{q}_{21})+m_{Q_1}} {(q_1 + q_{21})^2-m^2_{Q_1}} {\cal R}^{\nu} \Pi^{\alpha}_{q_2}(q) \gamma^\sigma \right]_{q=0}, \nonumber\\
{\cal A}^{\nu (S=0, L=0)}_2 &=& i Tr \left[\Pi^0_{q_1}(0) {\cal R}^{\nu}  \frac{-(\slashed{q}_2+ \slashed{q}_{12})+m_{\bar{Q}_2}} {(q_2 + q_{12})^2-m^2_{\bar{Q}_2}} \gamma^\sigma \Pi^{\alpha}_{q_2}(q) \gamma^\sigma \right]_{q=0}, \nonumber\\
{\cal A}^{\nu (S=0, L=0)}_3 &=& i Tr \left[\Pi^0_{q_1}(0) \gamma^\sigma \Pi^{\alpha}_{q_2}(q) {\cal R}^{\nu} \frac{-(\slashed{q}_1+ \slashed{q}_{22})+m_{\bar{Q}'_1}} {(q_1 + q_{22})^2-m^2_{\bar{Q}'_1}} \gamma^\sigma \right]_{q=0},\nonumber\\
{\cal A}^{\nu (S=0, L=0)}_4 &=& i Tr \left[\Pi^0_{q_1}(0) \gamma^\sigma \Pi^{\alpha}_{q_2}(q) \gamma^\sigma  \frac{\slashed{q}_2+ \slashed{q}_{11}+m_{Q'_2}} {(q_2 + q_{11})^2-m^2_{Q'_2}} {\cal R}^{\nu}  \right]_{q=0},  \nonumber\\
{\cal A}^{\nu (S=1, L=0)}_1 &=& i \epsilon_\alpha Tr \left[\Pi^{\alpha'}_{q_1}(0) \gamma^\sigma  \frac{(\slashed{q}_1+ \slashed{q}_{21})+m_{Q_1}} {(q_1 + q_{21})^2-m^2_{Q_1}} {\cal R}^{\nu} \Pi^{\alpha}_{q_2}(q) \gamma^\sigma \right]_{q=0}, \nonumber\\
{\cal A}^{\nu (S=1, L=0)}_2 &=& i \epsilon_\alpha Tr \left[\Pi^{\alpha'}_{q_1}(0) {\cal R}^{\nu}  \frac{-(\slashed{q}_2+ \slashed{q}_{12})+m_{\bar{Q}_2}} {(q_2 + q_{12})^2-m^2_{\bar{Q}_2}} \gamma^\sigma \Pi^{\alpha}_{q_2}(q) \gamma^\sigma \right]_{q=0}, \nonumber\\
{\cal A}^{\nu (S=1, L=0)}_3 &=& i \epsilon_\alpha Tr \left[\Pi^{\alpha'}_{q_1}(0) \gamma^\sigma \Pi^{\alpha}_{q_2}(q) {\cal R}^{\nu} \frac{-(\slashed{q}_1+ \slashed{q}_{22})+m_{\bar{Q}'_1}} {(q_1 + q_{22})^2-m^2_{\bar{Q}'_1}} \gamma^\sigma \right]_{q=0}, \nonumber\\
{\cal A}^{\nu (S=1, L=0)}_4 &=& i \epsilon_\alpha Tr \left[\Pi^{\alpha'}_{q_1}(0) \gamma^\sigma \Pi^{\alpha}_{q_2}(q) \gamma^\sigma  \frac{\slashed{q}_2+ \slashed{q}_{11}+m_{Q'_2}} {(q_2 + q_{11})^2-m^2_{Q'_2}} {\cal R}^{\nu}  \right]_{q=0}. \label{HadAmp-S}
\end{eqnarray}
\end{widetext}
Where $q$ being the relative momentum between the two constituent quarks of heavy quarkonium, the projectors $\Pi^{0}_{q_k}(q)$ spin-singlet states and $\Pi^{\alpha}_{q_k}(q)$ (with $k=1, 2$) spin-triplet states have the following form
\begin{eqnarray}
\Pi^{0}_{q_k}(q)&=&\frac{-\sqrt{m_{QQ'}}}{4 m_{Q_k}m_{Q'_k}}(\slashed{q}_{k2}- m_{Q'_k}) \gamma_5 (\slashed{q}_{k1} + m_{Q_k})\otimes  \frac{\delta_{ij}} {\sqrt{N_c}}, \nonumber\\
\Pi^{\alpha}_{q_k}(q)&=&\frac{-\sqrt{m_{QQ'}}}{4 m_{Q_k}m_{Q'_k}}(\slashed{q}_{k2}- m_{Q'_k}) \gamma_\alpha (\slashed{q}_{k1} + m_{Q_k})\otimes\frac{\delta_{ij}} {\sqrt{N_c}}.
\end{eqnarray}
where $q_{k1}=\frac{m_Q}{m_{Q\bar{Q'}}}{q_k}+q$ and $q_{k2}=\frac{m_{Q'}}{m_{Q\bar{Q'}}}{q_k}-q$ are the momenta of the two constituent heavy quarks, and $\delta_{ij}/\sqrt{N_c}$ is the color operator for color-singlet projector with $N_c=3$.
For the $S$-wave states, the relative momentum $q$ is set to zero directly.
The vertex $\cal{R}^\nu$ in Eq. (\ref{HadAmp-S}) is
\begin{equation}
    \cal{R}^\nu = \left\{
    \begin{array}{l}
        -i e e_{Q} \gamma^{\nu} \\
        \frac{-i g} {4 cos\theta_W} \gamma^\nu (1-4 e_Q  sin^2 \theta_W-\gamma^5)
    \end{array}
    \right.
\end{equation}
where the upper and lower expressions after the big left bracket are for the virtual photon and $Z^0$ propagated processes, respectively.
Here $e_Q=2/3$ for $c$-quark and $e_Q=-1/3$ for $b$-quark.
The Dirac $\gamma$ matrix chains ${\cal A}^\nu_k$ in Eq. (\ref{amplitude}) for the $P$-wave spin-singlet $^1P_1$ and the $P$-wave spin-triplet $^3P_J$ states ($J=0,1,2$) can be expressed in terms of the $^1S_0$-wave ones and the $^3S_1$-wave ones in Eq. (\ref{HadAmp-S}),
\begin{eqnarray}
{\cal A}^{\nu (S=0, L=1)}_k &=&  \epsilon_\beta(q_2) \left. \frac{d}{d q_\beta} {\cal A}^{\nu (S=0, L=0)}_k \right|_{q=0}, \nonumber\\
{\cal A}^{\nu (S=1, L=1)}_k &=&  \varepsilon^J_{\alpha \beta} (q_2)\left. \frac{d}{d q_\beta} {\cal A}^{\nu (S=1, L=0)}_k \right|_{q=0}. \label{HadAmp-P}
\end{eqnarray}
Where $k=1,2,3,4$, the first amplitudes are for $^1P_1$-wave spin-singlet states and the second are for $^3P_J$-wave spin-triplet states.
$\epsilon_\beta(q_2)$ is the polarization vector of the $^1P_1$ states, and $\varepsilon^{J}_{\alpha\beta}(q_2)$ is the polarization tensor for $^3P_J$ states($J=0,1,2$). The derivatives over the relative momentum $q_\beta$ in Eq. (\ref{HadAmp-P}) will be given complex and lengthy amplitudes.

When manipulating the squared amplitudes $|M(n)|^{2}$, the polarization vectors of the heavy quarkonium is needed to sum over.
For the spin-triplet $^3S_1$ states or the spin-singlet $^1P_1$ states, the polarization sum is given by \cite{nrqcd2}
\begin{eqnarray}\label{3pja}
\sum_{J_z}\epsilon_{\alpha} \epsilon_{\alpha'} = \Pi_{\alpha\alpha'} \equiv -g_{\alpha \alpha'}+\frac{q_{1\alpha}q_{1\alpha'}}{M_{Q\bar{Q'}}^{2}},
\end{eqnarray}
where $J_z=s_z$ or $l_z$ for $^3S_1$ and $1^1P_1$ states, respectively. In the case of $^3P_J$ states, the polarization sum should be performed by the selection of appropriate total angular momentum quantum number. The sum over polarization tensors  is given by \cite{nrqcd2}
\begin{eqnarray}\label{3pja}
\varepsilon^{(0)}_{\alpha\beta} \varepsilon^{(0)*}_{\alpha'\beta'} &=& \frac{1}{3} \Pi_{\alpha\beta}\Pi_{\alpha'\beta'}, \nonumber\\
\sum_{J_z}\varepsilon^{(1)}_{\alpha\beta} \varepsilon^{(1)*}_{\alpha'\beta'} &=& \frac{1}{2}
(\Pi_{\alpha\alpha'}\Pi_{\beta\beta'}- \Pi_{\alpha\beta'}\Pi_{\alpha'\beta}) \label{3pjb},\nonumber\\
\sum_{J_z}\varepsilon^{(2)}_{\alpha\beta} \varepsilon^{(2)*}_{\alpha'\beta'} &=& \frac{1}{2}
(\Pi_{\alpha\alpha'}\Pi_{\beta\beta'}+ \Pi_{\alpha\beta'}\Pi_{\alpha'\beta})-\frac{1}{3} \Pi_{\alpha\beta}\Pi_{\alpha'\beta'}, \nonumber\\ \label{3pjc}
\end{eqnarray}
for total angular momentum $J=0,1,2$, respectively.
To get compact analytical expression of the complicated $P$-wave channels and also improve the efficiency of numerical evaluation, the ``improved trace technology" is adopted to simplify the amplitudes ${\cal M}(n)$ at the amplitude level before evaluating the polarization sum.
To shorten this manuscript, we don't present it. For detailed techniques and more examples, one can refer to literatures \cite{cjx,lxz,Yang:2011ps,wbc1,Liao:2012rh,lx,Liao:2015vqa,lx1,lx2,qjpg}.

\section{Numerical Results}
\subsection{Input parameters}
\begin{table}
\caption{Masses (units: GeV) of the constituent quark and radial wave functions at the origin $|R_{|(Q\bar{Q^{'}})[S]\rangle}(0)|^2$ (units: GeV$^3$) and their first derivatives at the origin $|R'_{|(Q\bar{Q'})[P]\rangle}(0)|^2$ (units: GeV$^5$) within the BT-potential model~\cite{lx}.
Uncertainties of radial wave functions at the origin and their first derivatives at the origin
are caused by the corresponding varying quark masses.}
\begin{tabular}{|c|c|}
\hline\hline
$n_f=3~~~~m_c$,~~~$|R_{|(c\bar{c})[1S]\rangle}(0)|^2$~&~$m_c$,~~~$|R'_{|(c\bar{c})[1P]\rangle}(0)|^2$.\\
\hline
1.48$\pm$0.1,~~~$2.458^{+0.227}_{-0.327}$&1.75$\pm$0.1,~~~$0.322^{+0.077}_{-0.068}$\\
\hline\hline
$n_f=4~~~~m_b$,~~~$|R_{|(b\bar{b})[1S]\rangle}(0)|^2$~&~$m_b$,~~~$|R'_{|(b\bar{b})[1P]\rangle}(0)|^2$\\
\hline
4.71$\pm$0.2,~~~$16.12^{+1.28}_{-1.23}$&4.94$\pm$0.2,~~~$5.874^{+0.728}_{-0.675}$\\
\hline\hline
$n_f=3~~~~m_c~/m_b$,~~~$|R_{|(b\bar{c})[1S]\rangle}(0)|^2$&$m_c~/m_b$,~~~$|R'_{|(b\bar{c})[1P]\rangle}(0)|^2$\\
\hline
1.45$\pm$0.1~/4.85$\pm$0.2,$3.848^{+0.474}_{-0.453}$&1.75$\pm$0.1~/4.93$\pm$0.2,$0.518^{+0.123}_{-0.105}$\\
\hline\hline
\end{tabular}
\label{M&R}
\end{table}

In our numerical calculation, the quark mass $m_{Q_i}~/m_{\bar{Q}'_i}$ is set to be the mass of heavy quarkonium $M_{(Q\bar{Q'})_i}=m_{Q_i}+m_{\bar{Q}'_i}$, which ensures the gauge invariance of the hard scattering amplitude under the NRQCD framework.
In our previous work \cite{lx}, we calculate the radial wave functions at the origin $R_{|(Q\bar{Q'})[nS]\rangle}(0)$ and the first derivatives of radial wave functions at the origin $R'_{|(Q\bar{Q}')[nP]\rangle}(0)$ for heavy quarkonium $|(c\bar{c})[n]\rangle$, $|(b\bar{b})[n]\rangle$, and~$|(b\bar{c})[n]\rangle/|(c\bar{b})[n]\rangle$ under five different potential models ($n=1, 2, \cdots$), respectively.
In this work, the masses of $c$- and $b$- quarks, and the results of the Buchm\"{u}ller and Tye potential model (BT-potential) \cite{pot2,wgs} for the heavy quarkonium are presented in Table \ref{M&R}, where $n_f$ is the number of active flavor quarks. $n_f=3$ is for $|(c\bar{c})[n]\rangle$ charmonium and $|(c\bar{b})[n]\rangle$~/$|(b\bar{c})[n]\rangle$ mesons. $n_f=4$ is for $|(b\bar{b})[n]\rangle$ bottomonium. The uncertainties of the production cross section of double heavy quarkonium will be discussed from the radial wave functions at the origin and their derivatives at the origin under different potential models in Section \ref{uncertainty}. Since the non-perturbative matrix elements depend on the heavy quark masses. The uncertainties of radial wave functions at the origin and their first derivatives at the origin are caused by the corresponding varying quark masses in Table \ref{M&R}.
The renormalization scale is set to $m_{(c\bar{c})}$ and $m_{(b\bar{c})}$ for $|(c\bar{c})[n]\rangle$ and $|(b\bar{c})[n]\rangle$ quarkonium, which leads to the leading-order running coupling constant $\alpha_s=0.26$, and $m_{(b\bar{b})}$ for $|(b\bar{b})[n]\rangle$ quarkonium, which leads to $\alpha_s=0.18$.
Other parameters have the following values \cite{pdg}: the fine structure constant $\alpha=e^2/4\pi=1/130.9$, the mass of $Z^0$ boson $m_Z =91.1876$ GeV and its total decay width $\Gamma_{Z^0}=2.4952$ GeV, the Weinberg angle $\theta_W=\arcsin\sqrt{0.23119}$, the Fermi constant $G_F=\frac{\sqrt{2}g^2}{8m_W^2}=1.16639 \times 10^{-5}$ GeV$^{-2}$ with $m_W= 80.399$ GeV.

\subsection{Double heavy quarkonium production in $e^-e^+\to \gamma^*/Z^0\to|(Q\bar{Q'})[n]\rangle+|(Q'\bar{Q})[n']\rangle$}
\label{production}
\begin{table}
\caption{Cross sections (units:~$fb$) for the production of double quarkonium $|(c\bar{c})[n]\rangle$~/$|(c\bar{c})[n']\rangle$ in $e^+e^-$ annihilation via $\gamma^*~/Z^0$  at the center-of-mass energy $\sqrt{s}=91.1876$GeV within the BT-potential model ($n_f=3$)~\cite{lx}.}
\begin{tabular}{|c|c|c|}
\hline\hline
$\sigma{(e^-e^+\to \gamma^* \to|(c\bar{c})[^1S_0]\rangle+|(c\bar{c})[^3S_1]\rangle)}$&~3.269~$\times 10^{-6}$~\\
\hline
$\sigma{(e^-e^+\to \gamma^* \to|(c\bar{c})[^1S_0]\rangle+|(c\bar{c})[^1P_1]\rangle)}$&~1.051~$\times 10^{-5}$~\\
\hline
$\sigma{(e^-e^+\to \gamma^* \to|(c\bar{c})[^3S_1]\rangle+|(c\bar{c})[^1P_1]\rangle)}$&~8.821~$\times 10^{-8}$~\\
\hline
$\sigma{(e^-e^+\to \gamma^* \to|(c\bar{c})[^3S_1]\rangle+|(c\bar{c})[^3P_0]\rangle)}$&~1.084~$\times 10^{-5}$~\\
\hline
$\sigma{(e^-e^+\to \gamma^* \to|(c\bar{c})[^3S_1]\rangle+|(c\bar{c})[^3P_1]\rangle)}$&~2.142~$\times 10^{-5}$~\\
\hline
$\sigma{(e^-e^+\to \gamma^* \to|(c\bar{c})[^3S_1]\rangle+|(c\bar{c})[^3P_2]\rangle)}$&~4.264~$\times 10^{-5}$~\\
\hline
$\sigma{(e^-e^+\to Z^0 \to|(c\bar{c})[^1S_0]\rangle+|(c\bar{c})[^3S_1]\rangle)}$&~1.895~$\times 10^{-4}$~\\
\hline
$\sigma{(e^-e^+\to Z^0 \to|(c\bar{c})[^3S_1]\rangle+|(c\bar{c})[^3S_1]\rangle)}$&~6.416~$\times 10^{-4}$~\\
\hline
$\sigma{(e^-e^+\to Z^0 \to|(c\bar{c})[^1S_0]\rangle+|(c\bar{c})[^1P_1]\rangle)}$&~6.076~$\times 10^{-4}$~\\
\hline
$\sigma{(e^-e^+\to Z^0 \to|(c\bar{c})[^3S_1]\rangle+|(c\bar{c})[^1P_1]\rangle)}$&~4.540~$\times 10^{-3}$~\\
\hline
$\sigma{(e^-e^+\to  Z^0 \to|(c\bar{c})[^1S_0]\rangle+|(c\bar{c})[^3P_0]\rangle)}$&~1.370~$\times 10^{-3}$~\\
\hline
$\sigma{(e^-e^+\to Z^0 \to|(c\bar{c})[^1S_0]\rangle+|(c\bar{c})[^3P_1]\rangle)}$&~1.887~$\times 10^{-4}$~\\
\hline
$\sigma{(e^-e^+\to Z^0 \to|(c\bar{c})[^1S_0]\rangle+|(c\bar{c})[^3P_2]\rangle)}$&~2.842~$\times 10^{-3}$~\\
\hline
$\sigma{(e^-e^+\to Z^0 \to|(c\bar{c})[^3S_1]\rangle+|(c\bar{c})[^3P_0]\rangle)}$&~4.403~$\times 10^{-4}$~\\
\hline
$\sigma{(e^-e^+\to Z^0 \to|(c\bar{c})[^3S_1]\rangle+|(c\bar{c})[^3P_1]\rangle)}$&~1.762~$\times 10^{-4}$~\\
\hline
$\sigma{(e^-e^+\to Z^0 \to|(c\bar{c})[^3S_1]\rangle+|(c\bar{c})[^3P_2]\rangle)}$&~5.751~$\times 10^{-4}$~\\
\hline\hline
\end{tabular}
\label{tabrpa}
\end{table}

\begin{table}
\caption{Cross sections (units:~$fb$) for the production of double quarkonium $|(b\bar{b})[n]\rangle$~/$|(b\bar{b})[n']\rangle$ in $e^+e^-$ annihilation via $\gamma^*~/Z^0$ at the center-of-mass energy $\sqrt{s}=91.187$6 $GeV$ within the BT-potential model ($n_f=4$)~\cite{lx}.}
\begin{tabular}{|c|c|}
\hline\hline
$\sigma{(e^-e^+\to \gamma^* \to|(b\bar{b})[^1S_0]\rangle+|(b\bar{b})[^3S_1]\rangle)}$&~1.588~$\times 10^{-5}$~\\
\hline
$\sigma{(e^-e^+\to \gamma^* \to|(b\bar{b})[^1S_0]\rangle+|(b\bar{b})[^1P_1]\rangle)}$&~2.021~$\times 10^{-6}$~\\
\hline
$\sigma{(e^-e^+\to \gamma^* \to|(b\bar{b})[^3S_1]\rangle+|(b\bar{b})[^1P_1]\rangle)}$&~1.618~$\times 10^{-7}$~\\
\hline
$\sigma{(e^-e^+\to \gamma^* \to|(b\bar{b})[^3S_1]\rangle+|(b\bar{b})[^3P_0]\rangle)}$&~3.747~$\times 10^{-6}$~\\
\hline
$\sigma{(e^-e^+\to \gamma^* \to|(b\bar{b})[^3S_1]\rangle+|(b\bar{b})[^3P_1]\rangle)}$&~8.293~$\times 10^{-6}$~\\
\hline
$\sigma{(e^-e^+\to \gamma^* \to|(b\bar{b})[^3S_1]\rangle+|(b\bar{b})[^3P_2]\rangle)}$&~1.509~$\times 10^{-5}$~\\
\hline
$\sigma{(e^-e^+\to Z^0 \to|(b\bar{b})[^1S_0]\rangle+|(b\bar{b})[^3S_1]\rangle)}$&~4.286~$\times 10^{-2}$~\\
\hline
$\sigma{(e^-e^+\to Z^0 \to|(b\bar{b})[^3S_1]\rangle+|(b\bar{b})[^3S_1]\rangle)}$&~1.199~$\times 10^{-2}$~\\
\hline
$\sigma{(e^-e^+\to Z^0 \to|(b\bar{b})[^1S_0]\rangle+|(b\bar{b})[^1P_1]\rangle)}$&~5.327~$\times 10^{-3}$~\\
\hline
$\sigma{(e^-e^+\to Z^0 \to|(b\bar{b})[^3S_1]\rangle+|(b\bar{b})[^1P_1]\rangle)}$&~6.146~$\times 10^{-3}$~\\
\hline
$\sigma{(e^-e^+\to Z^0 \to|(b\bar{b})[^1S_0]\rangle+|(b\bar{b})[^3P_0]\rangle)}$&~9.215~$\times 10^{-4}$~\\
\hline
$\sigma{(e^-e^+\to Z^0 \to|(b\bar{b})[^1S_0]\rangle+|(b\bar{b})[^3P_1]\rangle)}$&~1.431~$\times 10^{-3}$~\\
\hline
$\sigma{(e^-e^+\to Z^0 \to|(b\bar{b})[^1S_0]\rangle+|(b\bar{b})[^3P_2]\rangle)}$&~2.815~$\times 10^{-3}$~\\
\hline
$\sigma{(e^-e^+\to Z^0 \to|(b\bar{b})[^3S_1]\rangle+|(b\bar{b})[^3P_0]\rangle)}$&~6.837~$\times 10^{-3}$~\\
\hline
$\sigma{(e^-e^+\to Z^0 \to|(b\bar{b})[^3S_1]\rangle+|(b\bar{b})[^3P_1]\rangle)}$&~1.175~$\times 10^{-2}$~\\
\hline
$\sigma{(e^-e^+\to Z^0 \to|(b\bar{b})[^3S_1]\rangle+|(b\bar{b})[^3P_2]\rangle)}$&~2.306~$\times 10^{-2}$~\\
\hline\hline
\end{tabular}
\label{tabrpb}
\end{table}

\begin{table}
\caption{Cross sections (units:~$fb$) for the production of double quarkonium $|(c\bar{b})[n]\rangle$~/$|(b\bar{c})[n']\rangle$ in $e^+e^-$ annihilation via $\gamma^*/Z^0$ at the center-of-mass energy $\sqrt{s}=91.187$6 $GeV$ within the BT-potential model ($n_f=3$)~\cite{lx}.}
\begin{tabular}{|c|c|}
\hline\hline
$\sigma{(e^-e^+\to \gamma^* \to|(c\bar{b})[^1S_0]\rangle+|(b\bar{c})[^3S_1]\rangle)}$&~5.711~$\times 10^{-5}$~\\
\hline
$\sigma{(e^-e^+\to \gamma^* \to|(c\bar{b})[^1S_0]\rangle+|(b\bar{c})[^1P_1]\rangle)}$&~2.288~$\times 10^{-6}$~\\
\hline
$\sigma{(e^-e^+\to \gamma^* \to|(c\bar{b})[^3S_1]\rangle+|(b\bar{c})[^1P_1]\rangle)}$&~1.746~$\times 10^{-6}$~\\
\hline
$\sigma{(e^-e^+\to \gamma^* \to|(c\bar{b})[^3S_1]\rangle+|(b\bar{c})[^3P_0]\rangle)}$&~1.039~$\times 10^{-5}$~\\
\hline
$\sigma{(e^-e^+\to \gamma^* \to|(c\bar{b})[^3S_1]\rangle+|(b\bar{c})[^3P_1]\rangle)}$&~8.443~$\times 10^{-5}$~\\
\hline
$\sigma{(e^-e^+\to \gamma^* \to|(c\bar{b})[^3S_1]\rangle+|(b\bar{c})[^3P_2]\rangle)}$&~2.816~$\times 10^{-4}$~\\
\hline
$\sigma{(e^-e^+\to Z^0 \to|(c\bar{b})[^1S_0]\rangle+|(b\bar{c})[^3S_1]\rangle)}$&~0.6346~~\\
\hline
$\sigma{(e^-e^+\to Z^0 \to|(c\bar{b})[^3S_1]\rangle+|(b\bar{c})[^3S_1]\rangle)}$&~1.150~\\
\hline
$\sigma{(e^-e^+\to Z^0 \to|(c\bar{b})[^1S_0]\rangle+|(b\bar{c})[^1P_1]\rangle)}$&~4.810~$\times 10^{-3}$~\\
\hline
$\sigma{(e^-e^+\to Z^0 \to|(c\bar{b})[^3S_1]\rangle+|(b\bar{c})[^1P_1]\rangle)}$&~1.177~$\times 10^{-2}$~\\
\hline
$\sigma{(e^-e^+\to Z^0 \to|(c\bar{b})[^1S_0]\rangle+|(b\bar{c})[^3P_0]\rangle)}$&~1.527~$\times 10^{-2}$~\\
\hline
$\sigma{(e^-e^+\to Z^0 \to|(c\bar{b})[^1S_0]\rangle+|(b\bar{c})[^3P_1]\rangle)}$&~6.162~$\times 10^{-3}$~\\
\hline
$\sigma{(e^-e^+\to Z^0 \to|(c\bar{b})[^1S_0]\rangle+|(b\bar{c})[^3P_2]\rangle)}$&~3.093~$\times 10^{-3}$~\\
\hline
$\sigma{(e^-e^+\to Z^0 \to|(c\bar{b})[^3S_1]\rangle+|(b\bar{c})[^3P_0]\rangle)}$&~0.4542~\\
\hline
$\sigma{(e^-e^+\to Z^0 \to|(c\bar{b})[^3S_1]\rangle+|(b\bar{c})[^3P_1]\rangle)}$&~0.2701~\\
\hline
$\sigma{(e^-e^+\to Z^0 \to|(c\bar{b})[^3S_1]\rangle+|(b\bar{c})[^3P_2]\rangle)}$&~1.133~~\\
\hline\hline
\end{tabular}
\label{tabrpc}
\end{table}


The total cross sections for the production of double heavy quarkonium via $e^-e^+\to \gamma^*/Z^0\to|(Q\bar{Q'})[n]\rangle+|(Q'\bar{Q})[n']\rangle$ ($Q/Q'=c$- or $b$- quarks) at center-of-momentum (CM) energy $\sqrt{s}=91.1876$ GeV are listed in Tables \ref{tabrpa}-\ref{tabrpc} for virtual photon $\gamma^*$ and $Z^0$ propagated processes, respectively. Here the BT-potential model is adopted to evaluate the non-perturbative hadronic matrix elements~\cite{lx} .
The cross sections for $1S$ and $1P$-wave charmonium in $e^-e^+\to \gamma^*/Z^0\to|(c\bar{c})[n]\rangle+|(c\bar{c})[n']\rangle$ at leading and next-to-leading order is calculated in Refs. \cite{gxz1,gxz2}. If the same input parameters are adopted, our estimations are consistent with theirs at leading order.

The total cross sections for the production of double heavy quarkonium via $e^-e^+\to \gamma^*/Z^0\to|(Q\bar{Q'})[n]\rangle+|(Q'\bar{Q})[n']\rangle$ ($Q/Q'=c \text{~or~} b$ quarks) at center-of-momentum (CM) energy $\sqrt{s}=91.1876$ GeV are summed for virtual photon $\gamma^*$ and $Z^0$ propagated processes at center-of-momentum (CM) energy $\sqrt{s}=91.1876$ GeV.
\begin{eqnarray}
\sigma{(e^-e^+\to \gamma^* \to|(c\bar{c})\rangle+|(c\bar{c})\rangle)}_{total} &=& 8.880\times 10^{-5}fb,\nonumber\\
\sigma{(e^-e^+\to Z^0 \to|(c\bar{c})\rangle+|(c\bar{c})\rangle)}_{total} &=& 1.158\times 10^{-2}fb,\nonumber\\
\sigma{(e^-e^+\to \gamma^* \to|(b\bar{b})\rangle+|(b\bar{b})\rangle)}_{total} &=&4.521\times 10^{-5}fb,\nonumber\\
\sigma{(e^-e^+\to Z^0 \to|(b\bar{b})\rangle+|(b\bar{b})\rangle)}_{total} &=& 0.113fb,\nonumber\\
\sigma{(e^-e^+\to \gamma^* \to|(c\bar{b})\rangle+|(b\bar{c})\rangle)} _{total} &=&4.37\times 10^{-4}fb,\nonumber\\
\sigma{(e^-e^+\to Z^0 \to|(c\bar{b})\rangle+|(b\bar{c})\rangle)}_{total} &=& 3.692fb.
\end{eqnarray}

For the $Z$ factory operation mode at CEPC, the designed integrated luminosity with two interaction point and in two years is $16~ab^{-1}$ \cite{CEPCStudyGroup:2018ghi}.
Then we can estimate the events of the production of double heavy quarkonium.
The double charmonium production are about 187 events for $e^-e^+\to \gamma^* / Z^0 \to|(c\bar{c})\rangle+|(c\bar{c})\rangle$, where the top three rankings for the double charmonium production are 73 events for $J/\psi+h_c$, 45 events for $\eta_c+\chi_{c2}$, and 22 events for $\eta_c+\chi_{c0}$.
The double bottomonium production are about 1808 events for $e^-e^+\to \gamma^* / Z^0 \to|(b\bar{b})\rangle+|(b\bar{b})\rangle$, where the top three rankings for the double bottomonium production are 686 events for $\eta_b+\Upsilon$, 369 events for $\Upsilon+\chi_{b2}$, and 192 events for $\Upsilon+\Upsilon$.
The double $B_c$ mesons production are about $5.907\times10^4$ events for $e^-e^+\to \gamma^* / Z^0 \to|(c\bar{b})\rangle+|(b\bar{c})\rangle$, where the top three rankings for the double $B_c$ production are $1.84\times10^4$ events for $B_c^*+B_c^*$, $1.81\times10^4$ events for $B_c^*+|(b\bar{c})[^3P_2]\rangle$, and $1.02\times10^4$ events for $B_c+B_c^*$.

\begin{figure*}[htbp]
\centering
\includegraphics[width=0.32\textwidth]{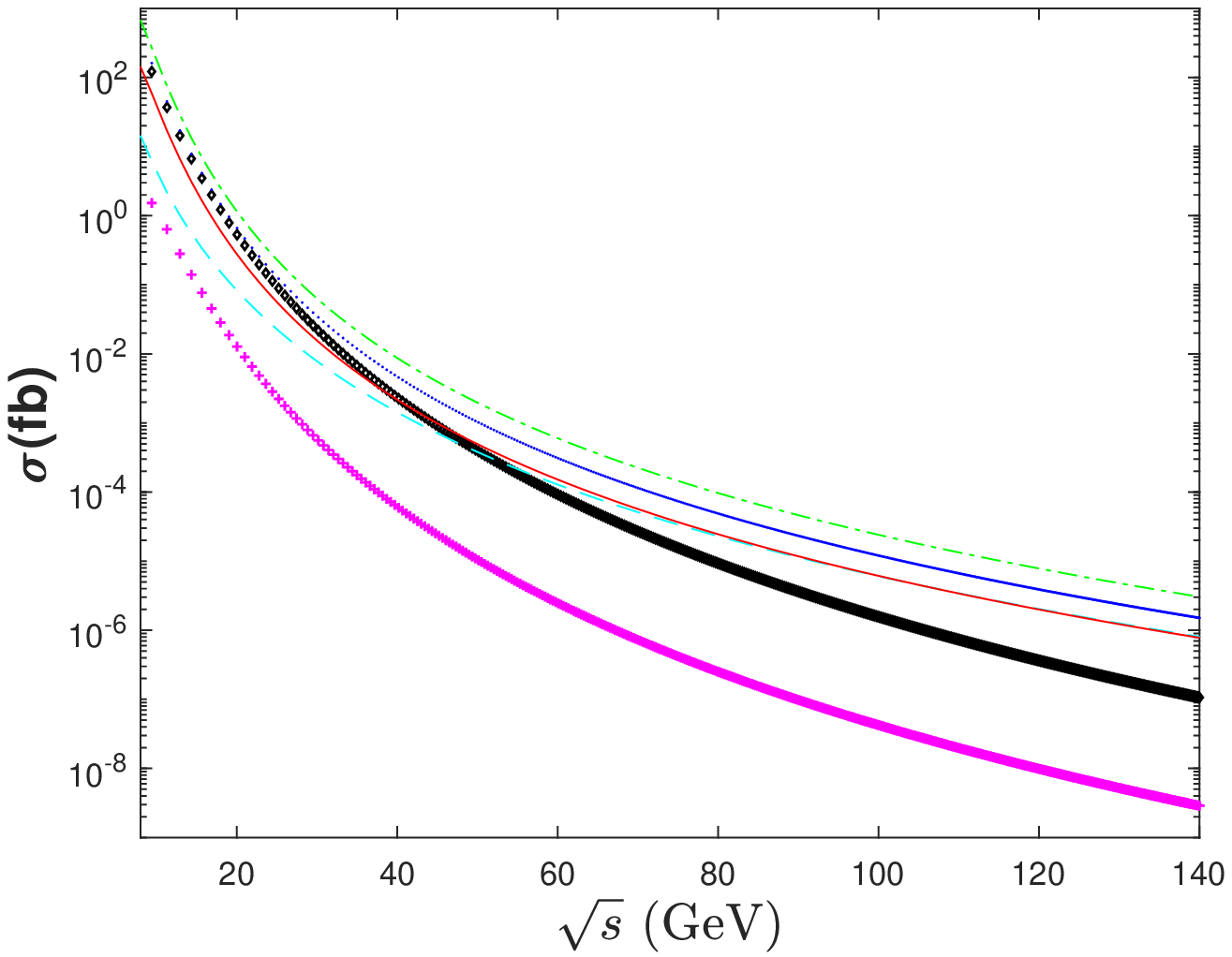}
\includegraphics[width=0.32\textwidth]{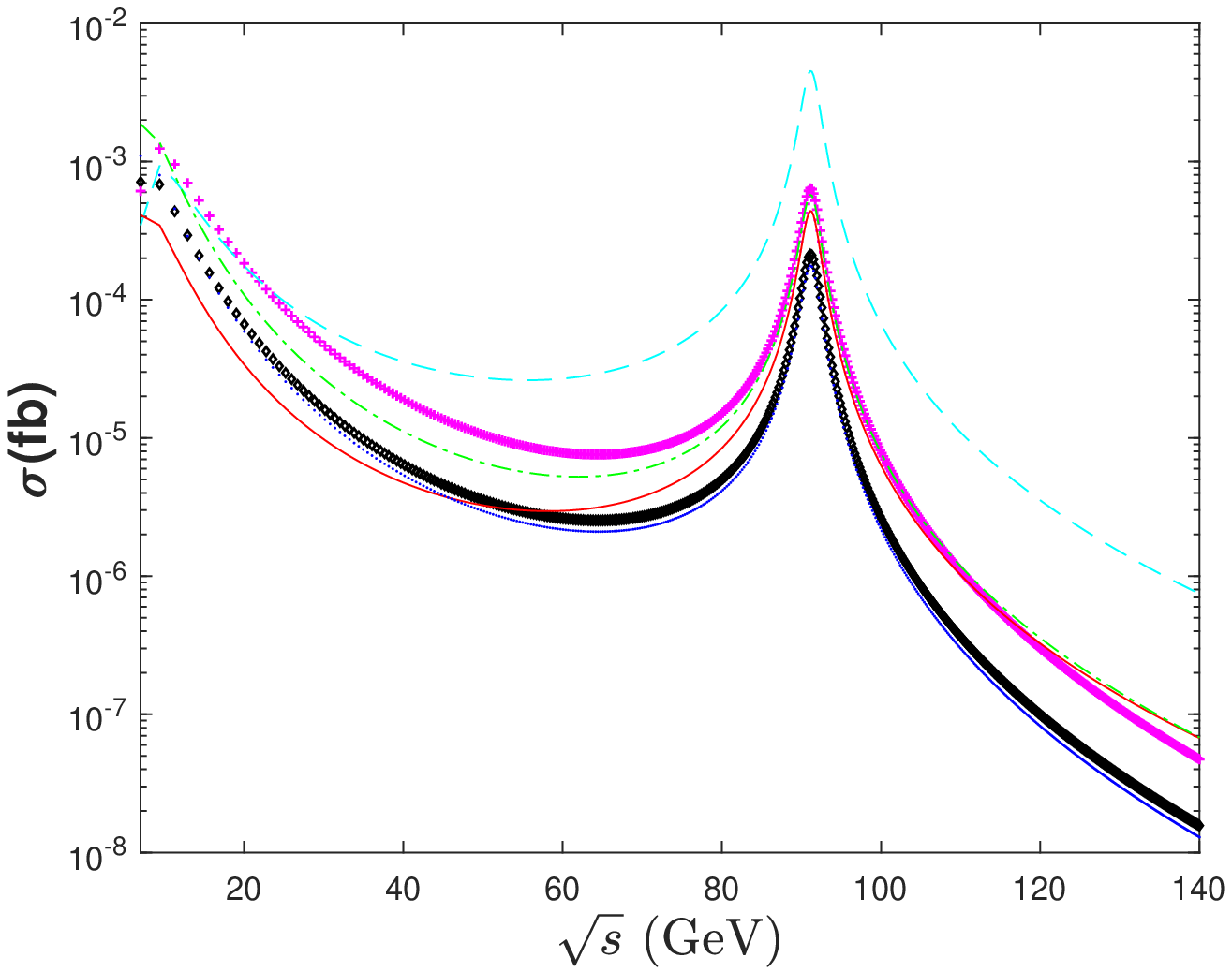}
\includegraphics[width=0.32\textwidth]{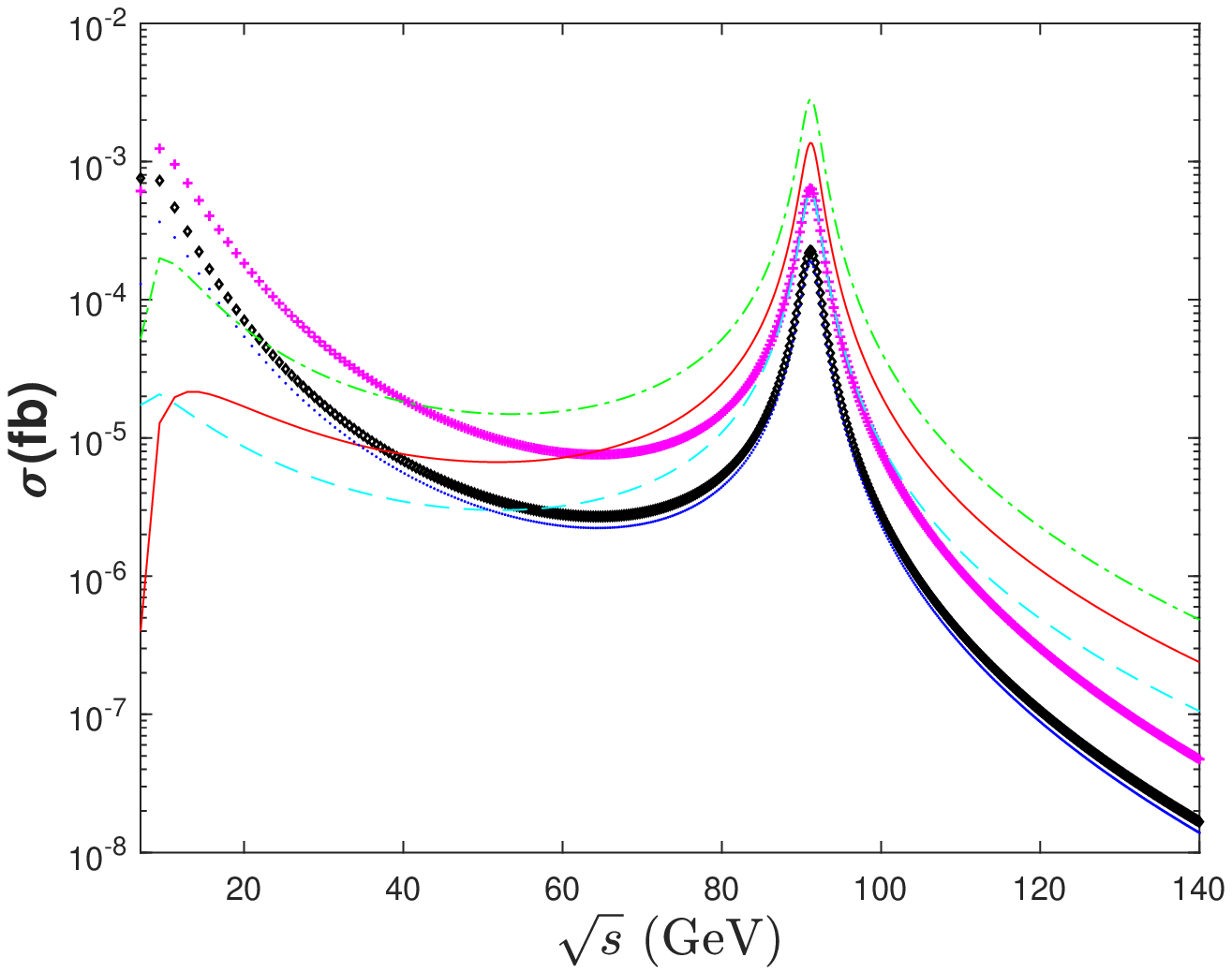}
\caption{
Cross sections versus the CM energy $\sqrt{s}$ for the channel $e^-e^+\to \gamma^*/Z^0 \to|(c\bar{c})[n]\rangle+|(c\bar{c})[n']\rangle$ via the virtual photon $\gamma^*$ (left), the $Z^0$ boson (middle) and (right). The diamond black line, cross magenta line, dashed cyan line, solid red line, dotted blue line, and the dash-dotted green line on the left are for the  double heavy quarkonium $|(c\bar{c})[^1S_0]\rangle/|(c\bar{c})[^3S_1]\rangle$, $|(c\bar{c})[^3S_1]\rangle/|(c\bar{c})[^1P_1]\rangle$,
$|(c\bar{c})[^1S_0]\rangle/|(c\bar{c})[^1P_1]\rangle$, $|(c\bar{c})[^3S_1]\rangle/|(c\bar{c})[^3P_0]\rangle$,
$|(c\bar{c})[^3S_1]\rangle/|(c\bar{c})[^3P_1]\rangle$, and $|(c\bar{c})[^3S_1]\rangle/|(c\bar{c})[^3P_2]\rangle$, respectively.
That on the middle are for the double heavy quarkonium $|(c\bar{c})[^1S_0]\rangle/|(c\bar{c})[^3S_1]\rangle$, $|(c\bar{c})[^3S_1]\rangle/|(c\bar{c})[^3S_1]\rangle$, $|(c\bar{c})[^3S_1]\rangle/|(c\bar{c})[^1P_1]\rangle$, $|(c\bar{c})[^3S_1]\rangle/|(c\bar{c})[^3P_0]\rangle$,
$|(c\bar{c})[^3S_1]\rangle/|(c\bar{c})[^3P_1]\rangle$, and $|(c\bar{c})[^3S_1]\rangle/|(c\bar{c})[^3P_2]\rangle$, respectively.
That on the right are for the double heavy quarkonium $|(c\bar{c})[^1S_0]\rangle/|(c\bar{c})[^3S_1]\rangle$, $|(c\bar{c})[^3S_1]\rangle/|(c\bar{c})[^3S_1]\rangle$, $|(c\bar{c})[^1S_0]\rangle/|(c\bar{c})[^1P_1]\rangle$, $|(c\bar{c})[^1S_0]\rangle/|(c\bar{c})[^3P_0]\rangle$,
$|(c\bar{c})[^1S_0]\rangle/|(c\bar{c})[^3P_1]\rangle$, and $|(c\bar{c})[^1S_0]\rangle/|(c\bar{c})[^3P_2]\rangle$, respectively.
} \label{ccds}
\end{figure*}

\begin{figure*}[htbp]
\centering
\includegraphics[width=0.32\textwidth]{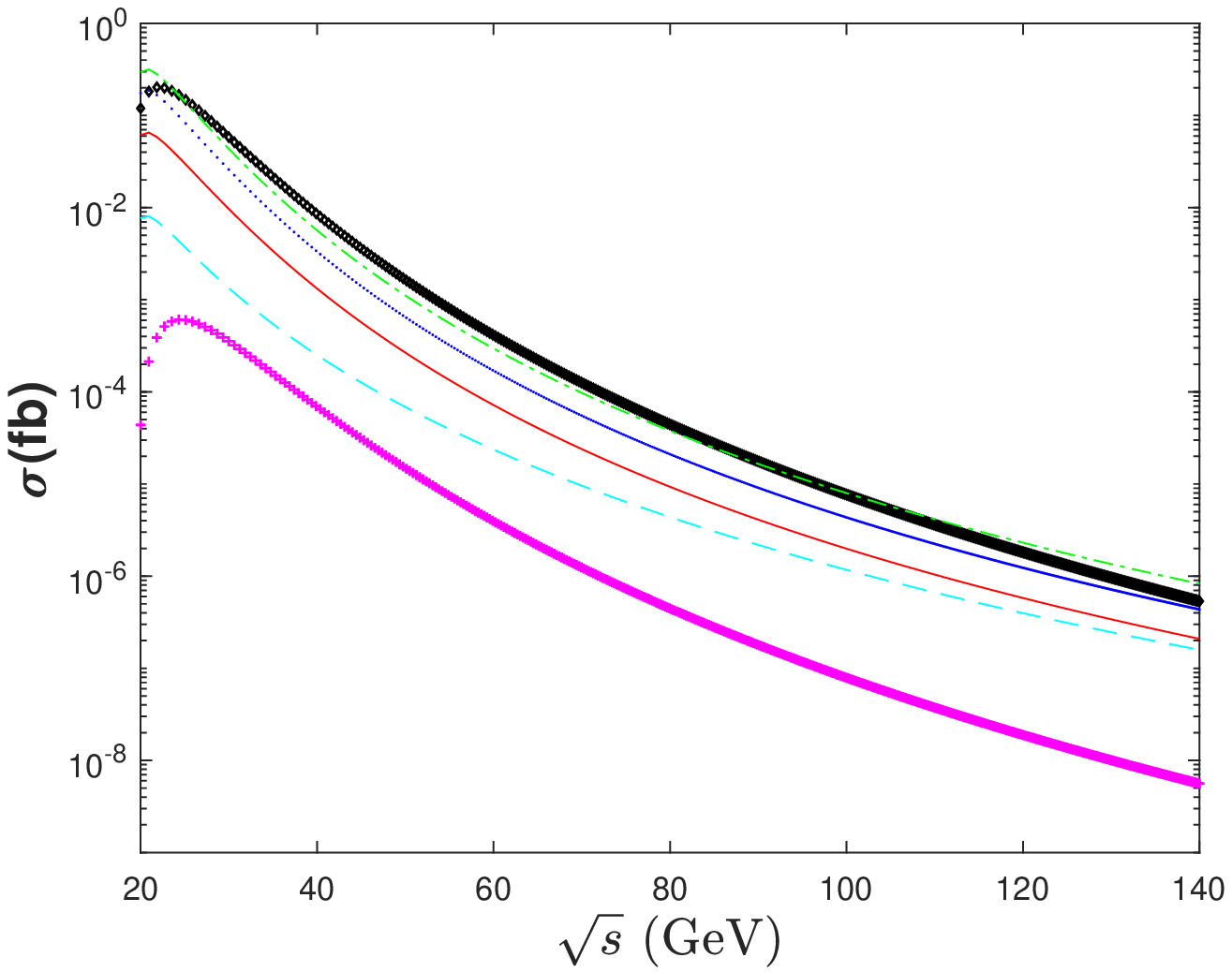}
\includegraphics[width=0.32\textwidth]{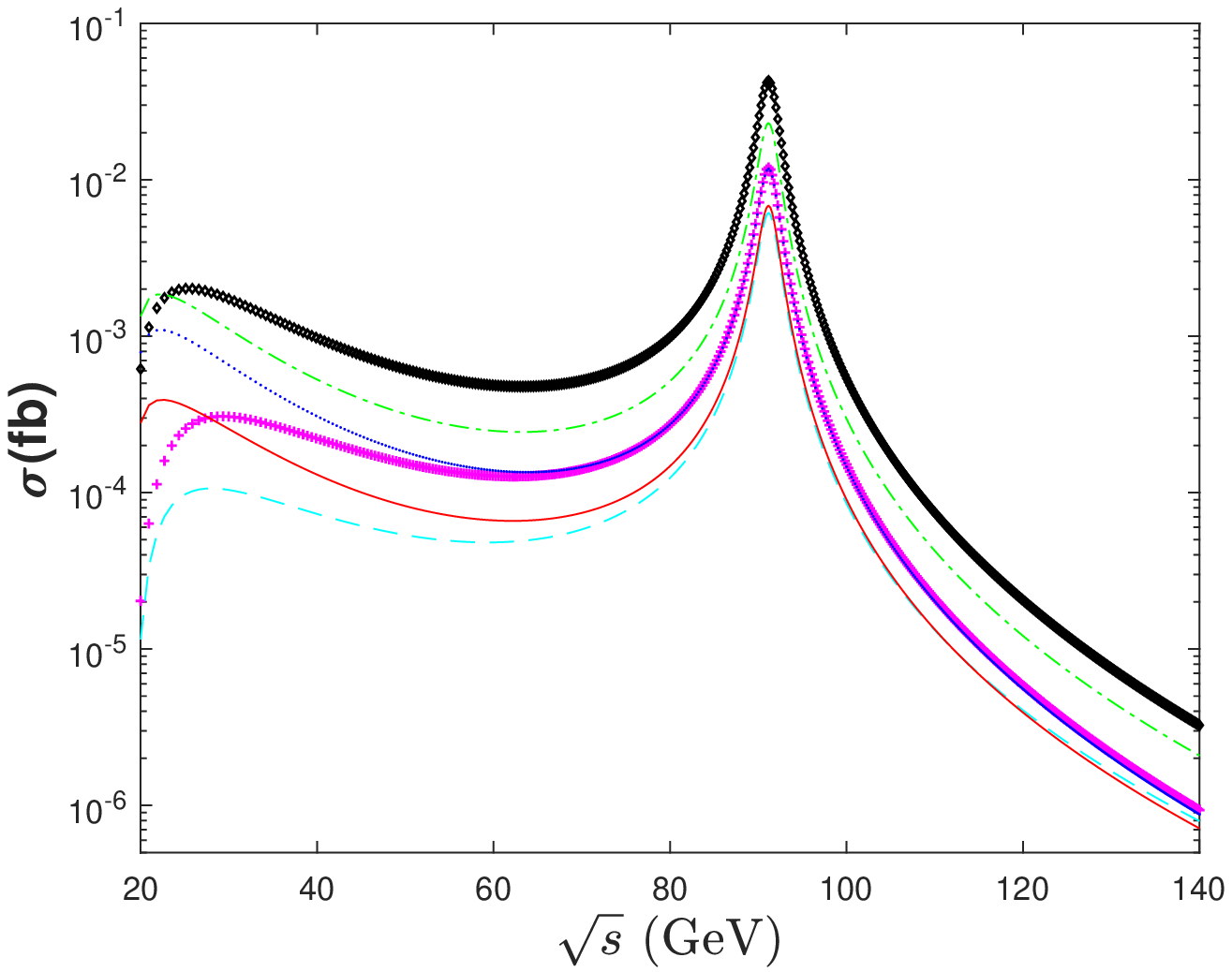}
\includegraphics[width=0.32\textwidth]{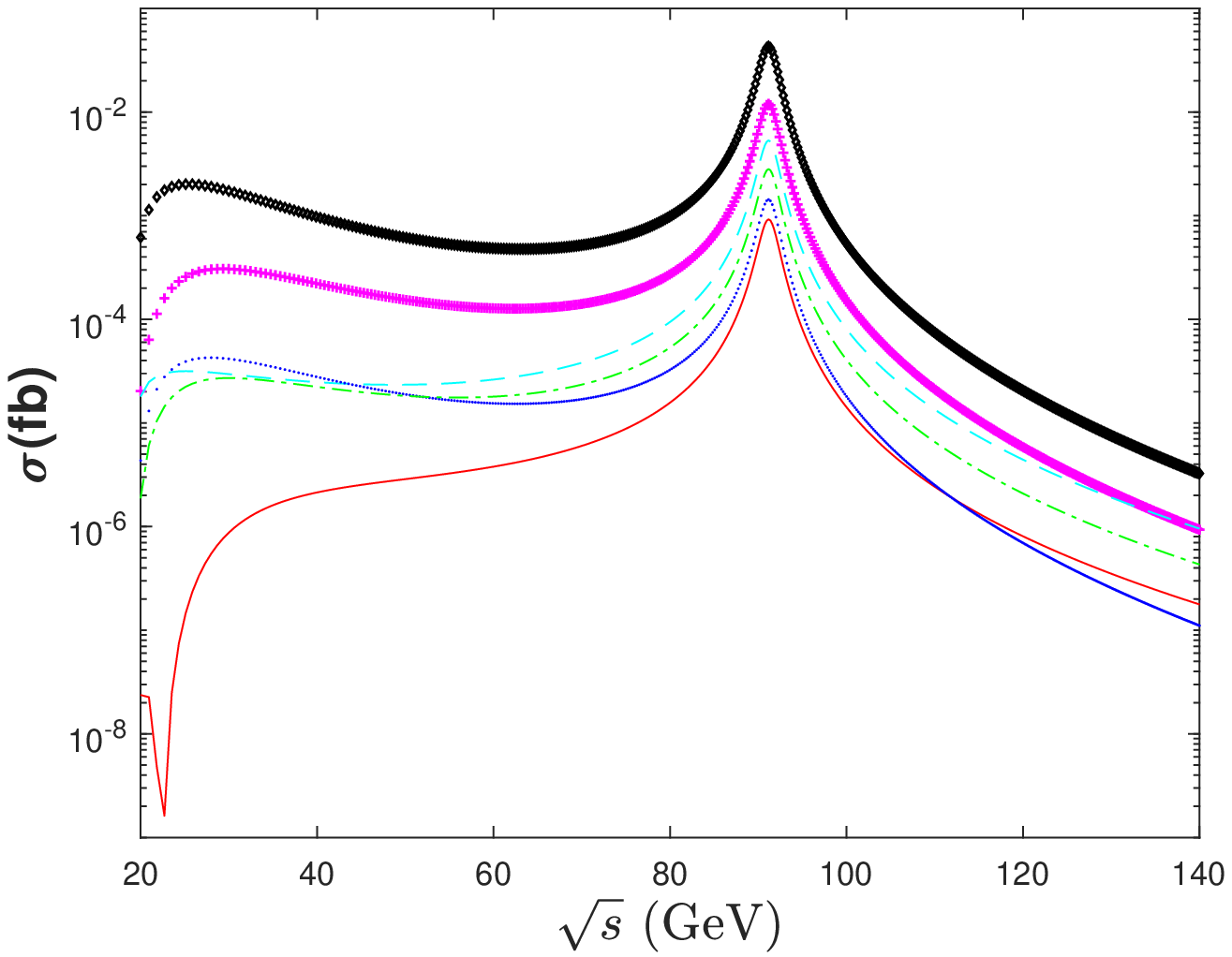}
\caption{
Cross sections versus the CM energy $\sqrt{s}$ for the channel $e^-e^+\to \gamma^*/Z^0 \to|(b\bar{b})[n]\rangle+|(b\bar{b})[n']\rangle$ via the virtual photon $\gamma^*$ (left), the $Z^0$ boson (middle) and (right). The diamond black line, cross magenta line, dashed cyan line, solid red line, dotted blue line, and the dash-dotted green line on the left are for the double heavy quarkonium $|(b\bar{b})[^1S_0]\rangle/|(b\bar{b})[^3S_1]\rangle$, $|(b\bar{b})[^3S_1]\rangle/|(b\bar{b})[^1P_1]\rangle$,
$|(b\bar{b})[^1S_0]\rangle/|(b\bar{b})[^1P_1]\rangle$, $|(b\bar{b})[^3S_1]\rangle/|(b\bar{b})[^3P_0]\rangle$,
$|(b\bar{b})[^3S_1]\rangle/|(b\bar{b})[^3P_1]\rangle$, and $|(b\bar{b})[^3S_1]\rangle/|(b\bar{b})[^3P_2]\rangle$, respectively.
That on the middle are for the double heavy quarkonium $|(b\bar{b})[^1S_0]\rangle/|(b\bar{b})[^3S_1]\rangle$, $|(b\bar{b})[^3S_1]\rangle/|(b\bar{b})[^3S_1]\rangle$, $|(b\bar{b})[^3S_1]\rangle/|(b\bar{b})[^1P_1]\rangle$, $|(b\bar{b})[^3S_1]\rangle/|(b\bar{b})[^3P_0]\rangle$,
$|(b\bar{b})[^3S_1]\rangle/|(b\bar{b})[^3P_1]\rangle$, and $|(b\bar{b})[^3S_1]\rangle/|(b\bar{b})[^3P_2]\rangle$, respectively.
That on the right are for the double heavy quarkonium $|(b\bar{b})[^1S_0]\rangle/|(b\bar{b})[^3S_1]\rangle$, $|(b\bar{b})[^3S_1]\rangle/|(b\bar{b})[^3S_1]\rangle$, $|(b\bar{b})[^1S_0]\rangle/|(b\bar{b})[^1P_1]\rangle$, $|(b\bar{b})[^1S_0]\rangle/|(b\bar{b})[^3P_0]\rangle$,
$|(b\bar{b})[^1S_0]\rangle/|(b\bar{b})[^3P_1]\rangle$, and $|(b\bar{b})[^1S_0]\rangle/|(b\bar{b})[^3P_2]\rangle$, respectively.
} \label{bbds}
\end{figure*}

\begin{figure*}[htbp]
\centering
\includegraphics[width=0.32\textwidth]{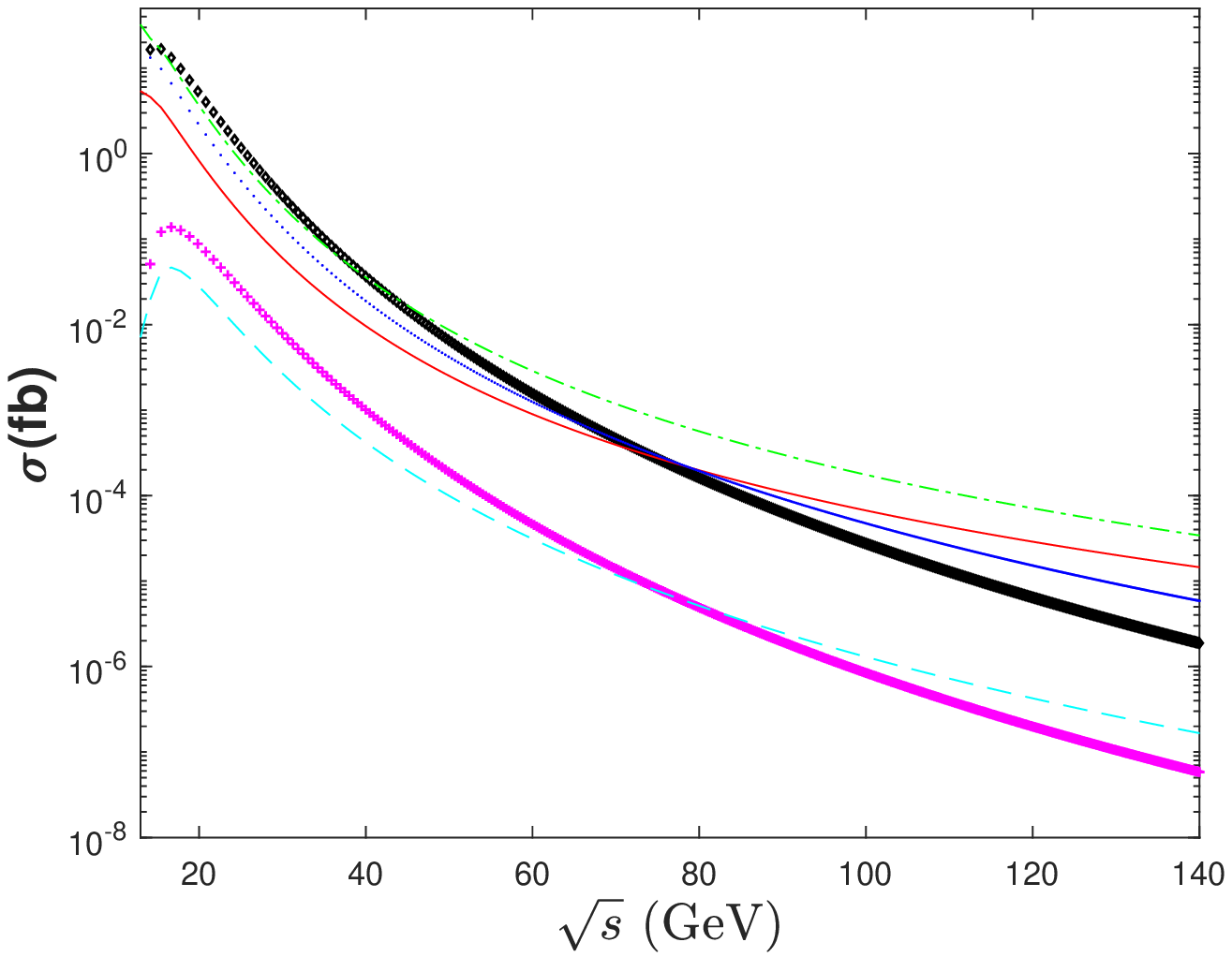}
\includegraphics[width=0.32\textwidth]{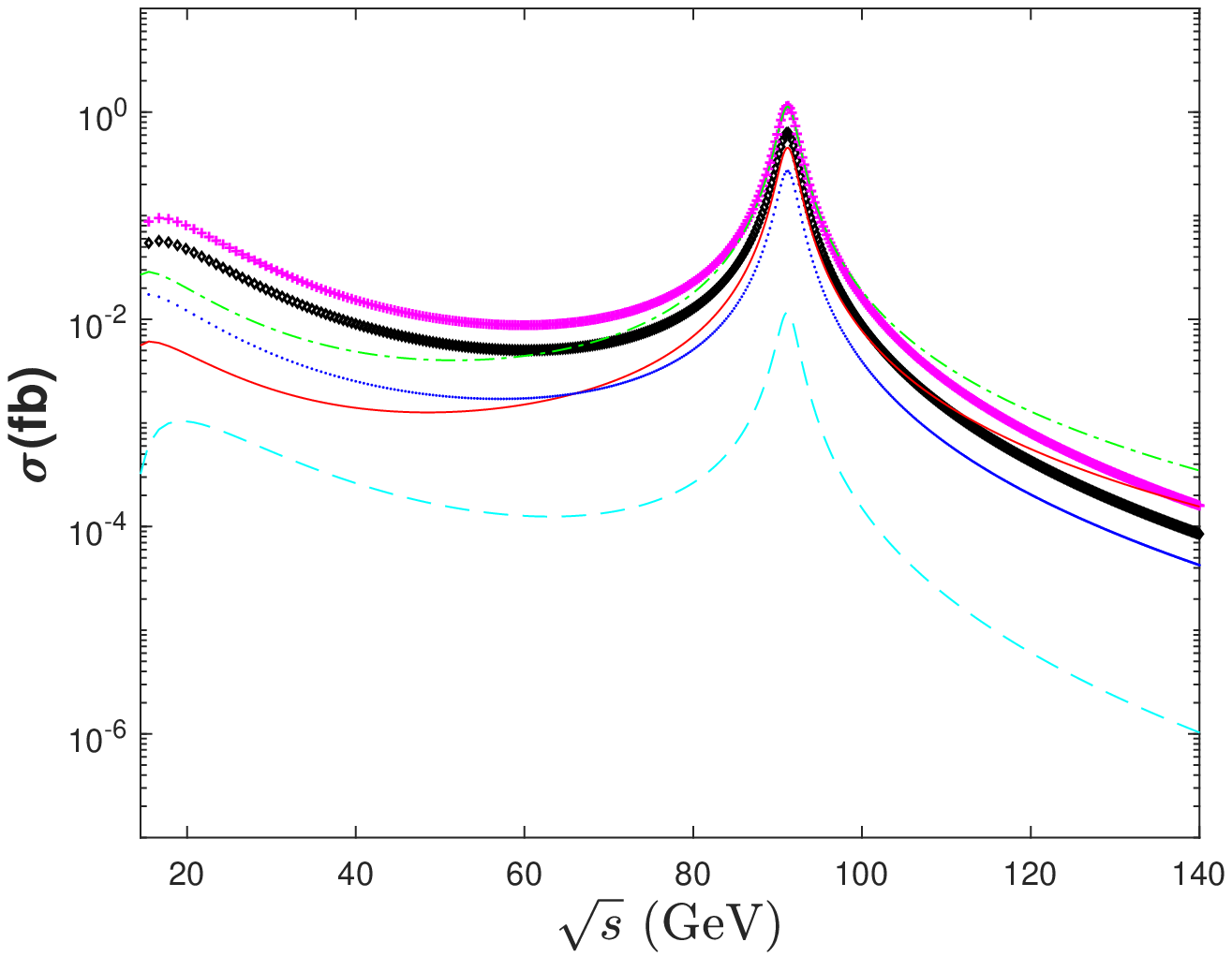}
\includegraphics[width=0.32\textwidth]{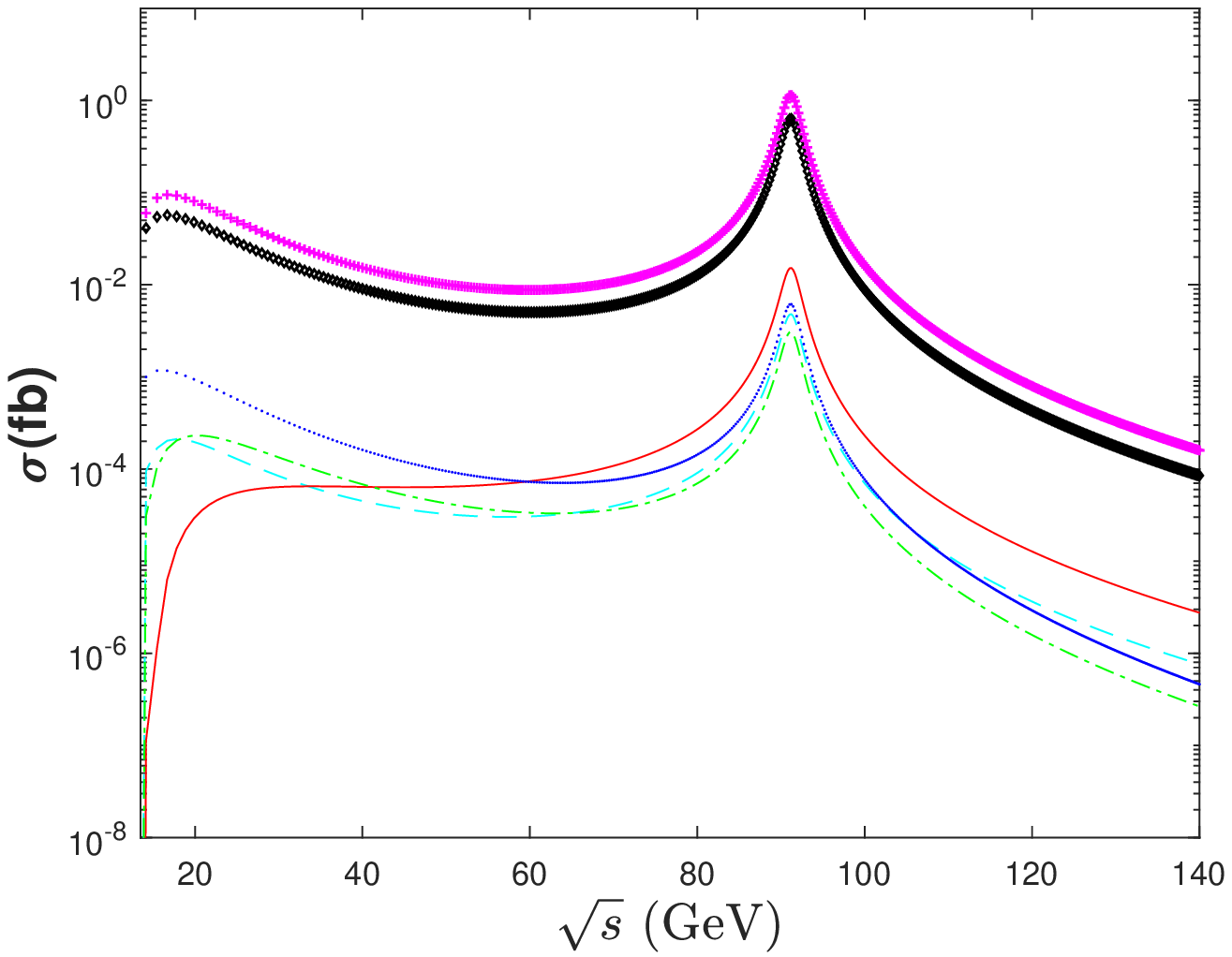}
\caption{
Cross sections versus the CM energy $\sqrt{s}$ for the channel $e^-e^+\to \gamma^*/Z^0 \to|(c\bar{b})[n]\rangle+|(b\bar{c})[n']\rangle$ via the virtual photon $\gamma^*$ (left), the $Z^0$ boson (middle) and (right). The diamond black line, cross magenta line, dashed cyan line, solid red line, dotted blue line, and the dash-dotted green line on the left are for the double heavy quarkonium $|(c\bar{b})[^1S_0]\rangle/|(b\bar{c})[^3S_1]\rangle$, $|(c\bar{b})[^3S_1]\rangle/|(b\bar{c})[^1P_1]\rangle$,
$|(c\bar{b})[^1S_0]\rangle/|(b\bar{c})[^1P_1]\rangle$, $|(c\bar{b})[^3S_1]\rangle/|(b\bar{c})[^3P_0]\rangle$,
$|(c\bar{b})[^3S_1]\rangle/|(b\bar{c})[^3P_1]\rangle$, and $|(c\bar{b})[^3S_1]\rangle/|(b\bar{c})[^3P_2]\rangle$, respectively.
That on the middle are for the double heavy quarkonium $|(c\bar{b})[^1S_0]\rangle/|(b\bar{c})[^3S_1]\rangle$, $|(c\bar{b})[^3S_1]\rangle/|(b\bar{c})[^3S_1]\rangle$, $|(c\bar{b})[^3S_1]\rangle/|(b\bar{c})[^1P_1]\rangle$, $|(c\bar{b})[^3S_1]\rangle/|(b\bar{c})[^3P_0]\rangle$,
$|(c\bar{b})[^3S_1]\rangle/|(b\bar{c})[^3P_1]\rangle$, and $|(c\bar{b})[^3S_1]\rangle/|(b\bar{c})[^3P_2]\rangle$, respectively.
That on the right are for the double heavy quarkonium $|(c\bar{b})[^1S_0]\rangle/|(b\bar{c})[^3S_1]\rangle$, $|(c\bar{b})[^3S_1]\rangle/|(b\bar{c})[^3S_1]\rangle$, $|(c\bar{b})[^1S_0]\rangle/|(b\bar{c})[^1P_1]\rangle$, $|(c\bar{b})[^1S_0]\rangle/|(b\bar{c})[^3P_0]\rangle$,
$|(c\bar{b})[^1S_0]\rangle/|(b\bar{c})[^3P_1]\rangle$, and $|(c\bar{b})[^1S_0]\rangle/|(b\bar{c})[^3P_2]\rangle$, respectively.
} \label{bcds}
\end{figure*}

In Figs. \ref{ccds}-\ref{bcds}, we display the total cross sections versus the CM energy $\sqrt{s}$ for double heavy quarkonium $|(c\bar{c})[n]\rangle$~/$|(c\bar{c})[n']\rangle$, $|(b\bar{b})[n]\rangle$~/$|(b\bar{b})[n']\rangle$, and $|(c\bar{b})[n]\rangle/|(b\bar{c})[n']\rangle$, respectively, where $[n]$~/$[n']$ stands for $^1S_0$,  $^3S_1$,  $^1P_1$, and $^3P_J$ quarkonium ($J=0,1,2$).
They show explicitly the contributions of $\gamma^*$ and $Z^0$ propagated processes from $\sqrt{s}=10$ GeV to 140 GeV.
Around the $Z^0$ peak, the $Z^0$ propagated processes dominate without any doubts.
\begin{figure*}[htbp]
\centering
\includegraphics[width=0.32\textwidth]{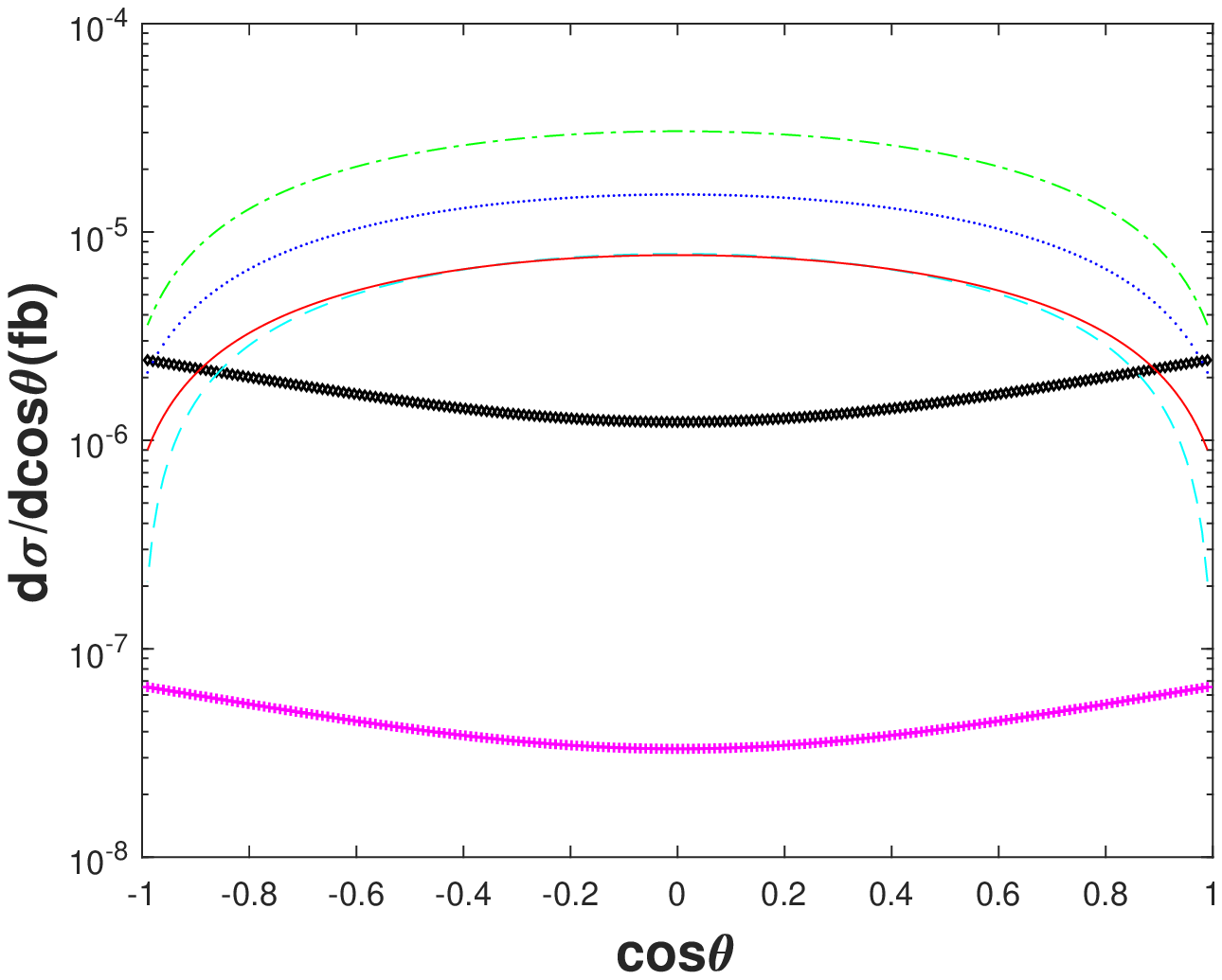}
\includegraphics[width=0.32\textwidth]{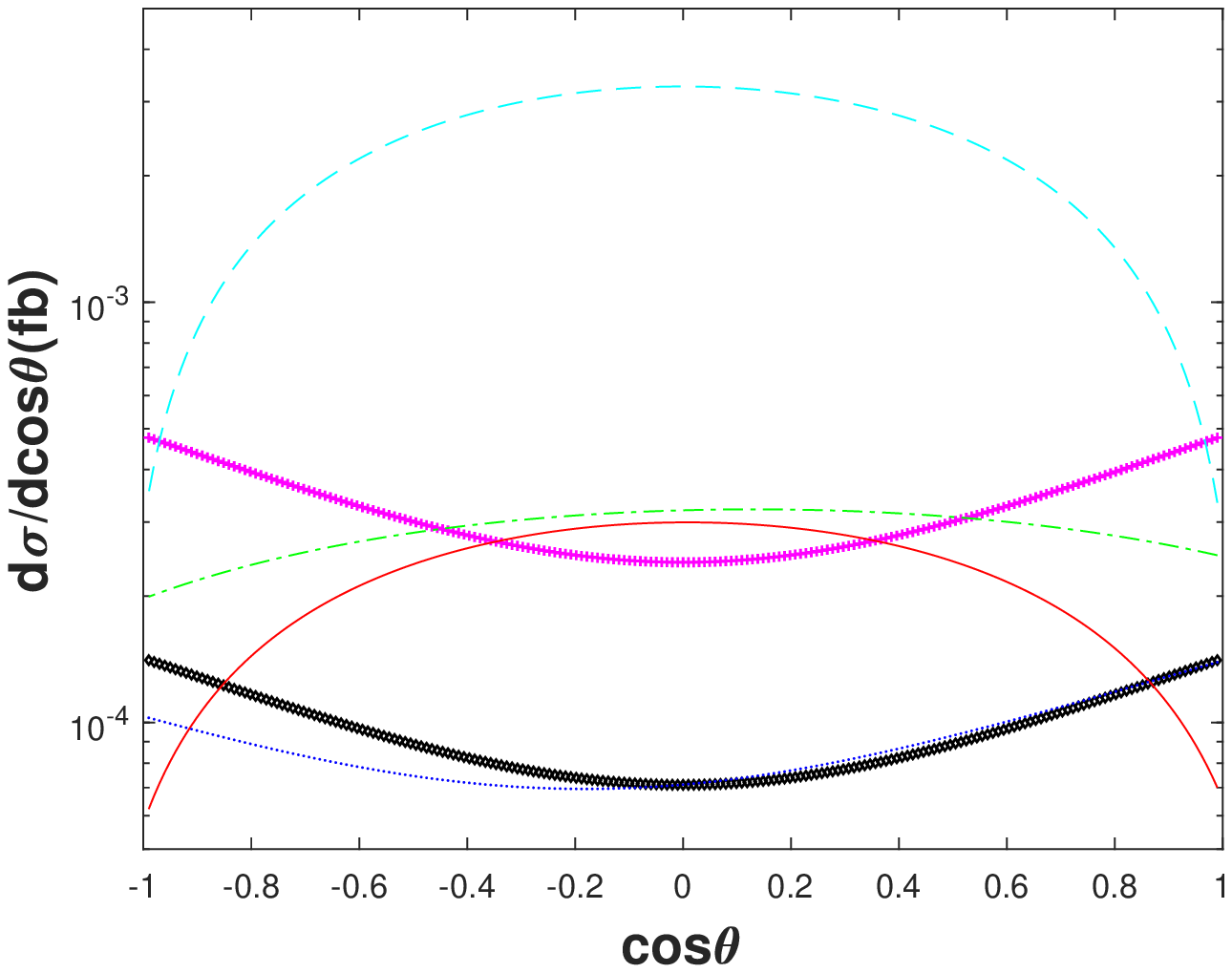}
\includegraphics[width=0.32\textwidth]{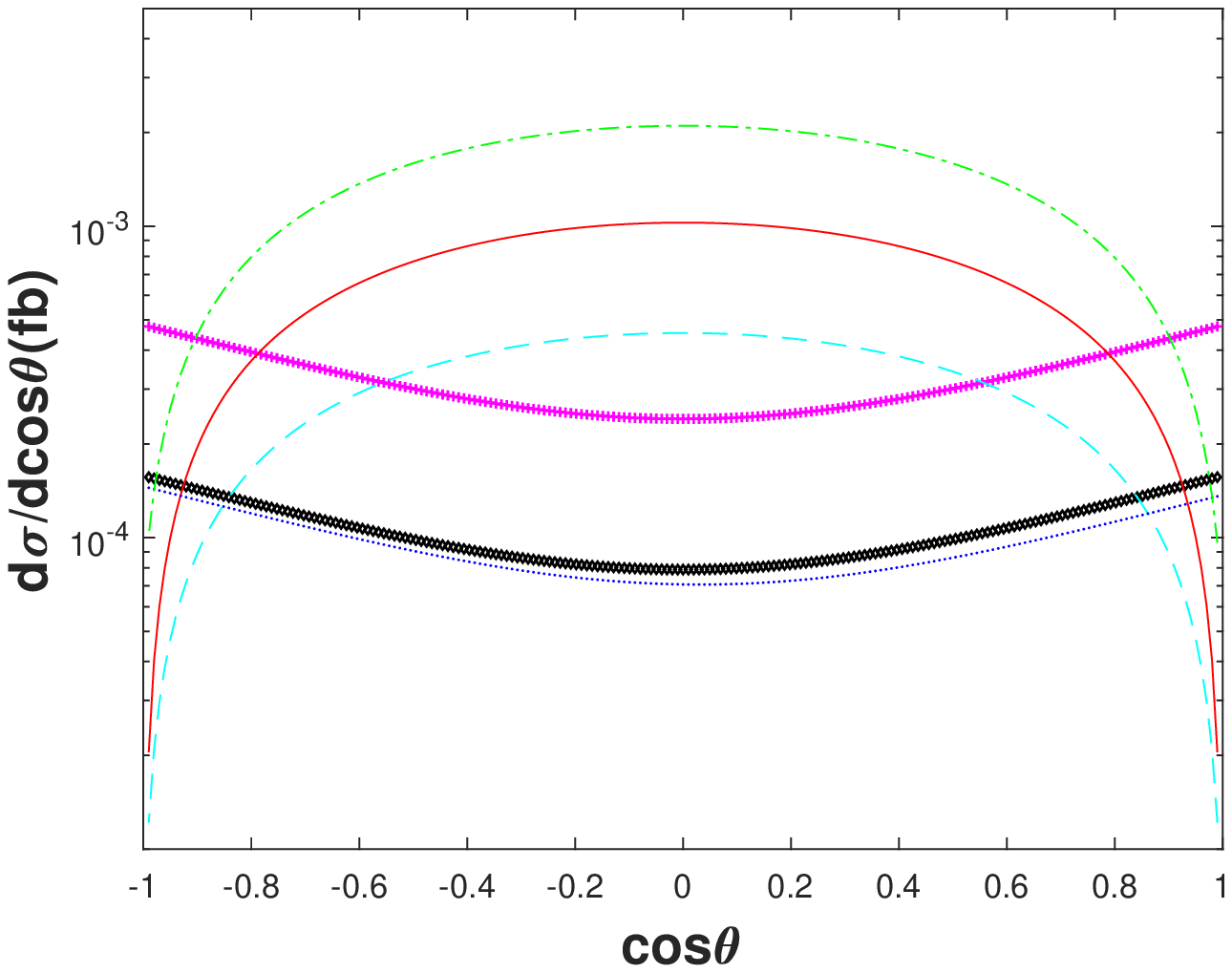}
\caption{
Differential cross sections d$\sigma$/dcos$\theta$ for the channel $e^-e^+\to \gamma^*/Z^0 \to|(c\bar{c})[n]\rangle+|(c\bar{c})[n']\rangle$ via the virtual photon $\gamma^*$ (left), the $Z^0$ boson (middle) and (right). The diamond black line, cross magenta line, dashed cyan line, solid red line, dotted blue line, and the dash-dotted green line on the left are for the double heavy quarkonium $|(c\bar{c})[^1S_0]\rangle/|(c\bar{c})[^3S_1]\rangle$, $|(c\bar{c})[^3S_1]\rangle/|(c\bar{c})[^1P_1]\rangle$,
$|(c\bar{c})[^1S_0]\rangle/|(c\bar{c})[^1P_1]\rangle$, $|(c\bar{c})[^3S_1]\rangle/|(c\bar{c})[^3P_0]\rangle$,
$|(c\bar{c})[^3S_1]\rangle/|(c\bar{c})[^3P_1]\rangle$, and $|(c\bar{c})[^3S_1]\rangle/|(c\bar{c})[^3P_2]\rangle$, respectively.
That on the middle are for the double heavy quarkonium $|(c\bar{c})[^1S_0]\rangle/|(c\bar{c})[^3S_1]\rangle$, $|(c\bar{c})[^3S_1]\rangle/|(c\bar{c})[^3S_1]\rangle$, $|(c\bar{c})[^3S_1]\rangle/|(c\bar{c})[^1P_1]\rangle$, $|(c\bar{c})[^3S_1]\rangle/|(c\bar{c})[^3P_0]\rangle$,
$|(c\bar{c})[^3S_1]\rangle/|(c\bar{c})[^3P_1]\rangle$, and $|(c\bar{c})[^3S_1]\rangle/|(c\bar{c})[^3P_2]\rangle$, respectively.
That on the right are for the double heavy quarkonium $|(c\bar{c})[^1S_0]\rangle/|(c\bar{c})[^3S_1]\rangle$, $|(c\bar{c})[^3S_1]\rangle/|(c\bar{c})[^3S_1]\rangle$, $|(c\bar{c})[^1S_0]\rangle/|(c\bar{c})[^1P_1]\rangle$, $|(c\bar{c})[^1S_0]\rangle/|(c\bar{c})[^3P_0]\rangle$,
$|(c\bar{c})[^1S_0]\rangle/|(c\bar{c})[^3P_1]\rangle$, and $|(c\bar{c})[^1S_0]\rangle/|(c\bar{c})[^3P_2]\rangle$, respectively.
} \label{ccdcos}
\end{figure*}

\begin{figure*}[htbp]
\centering
\includegraphics[width=0.32\textwidth]{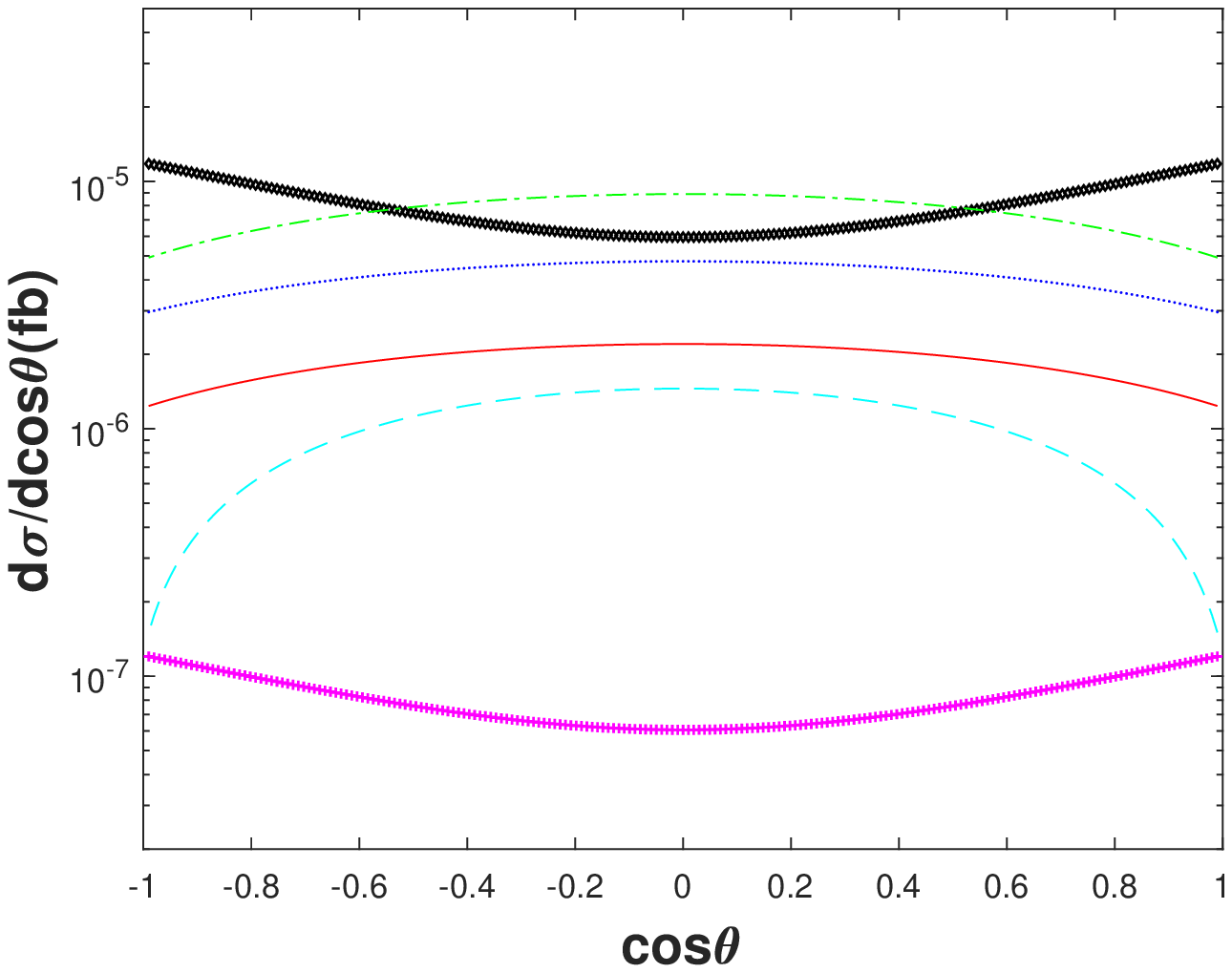}
\includegraphics[width=0.32\textwidth]{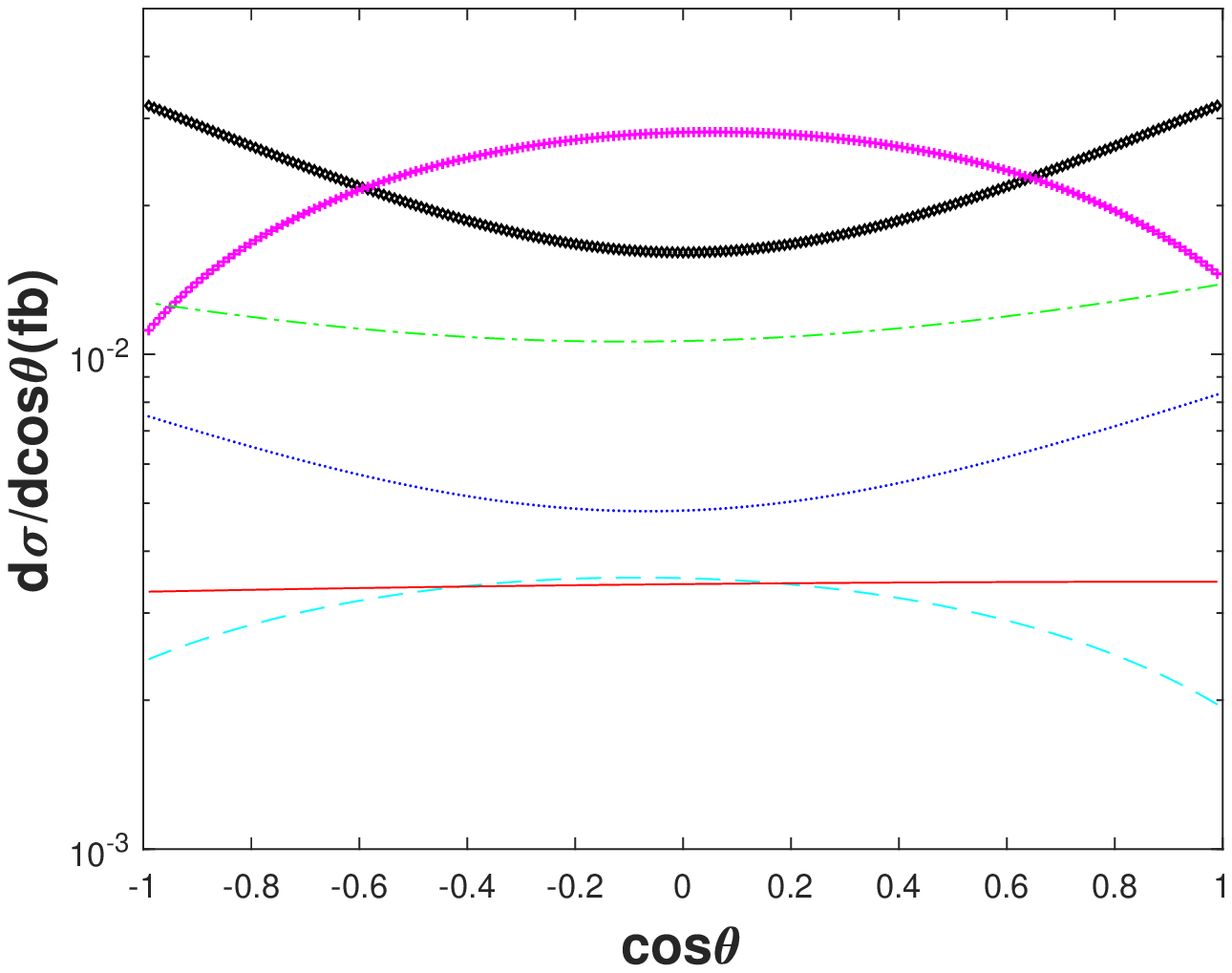}
\includegraphics[width=0.32\textwidth]{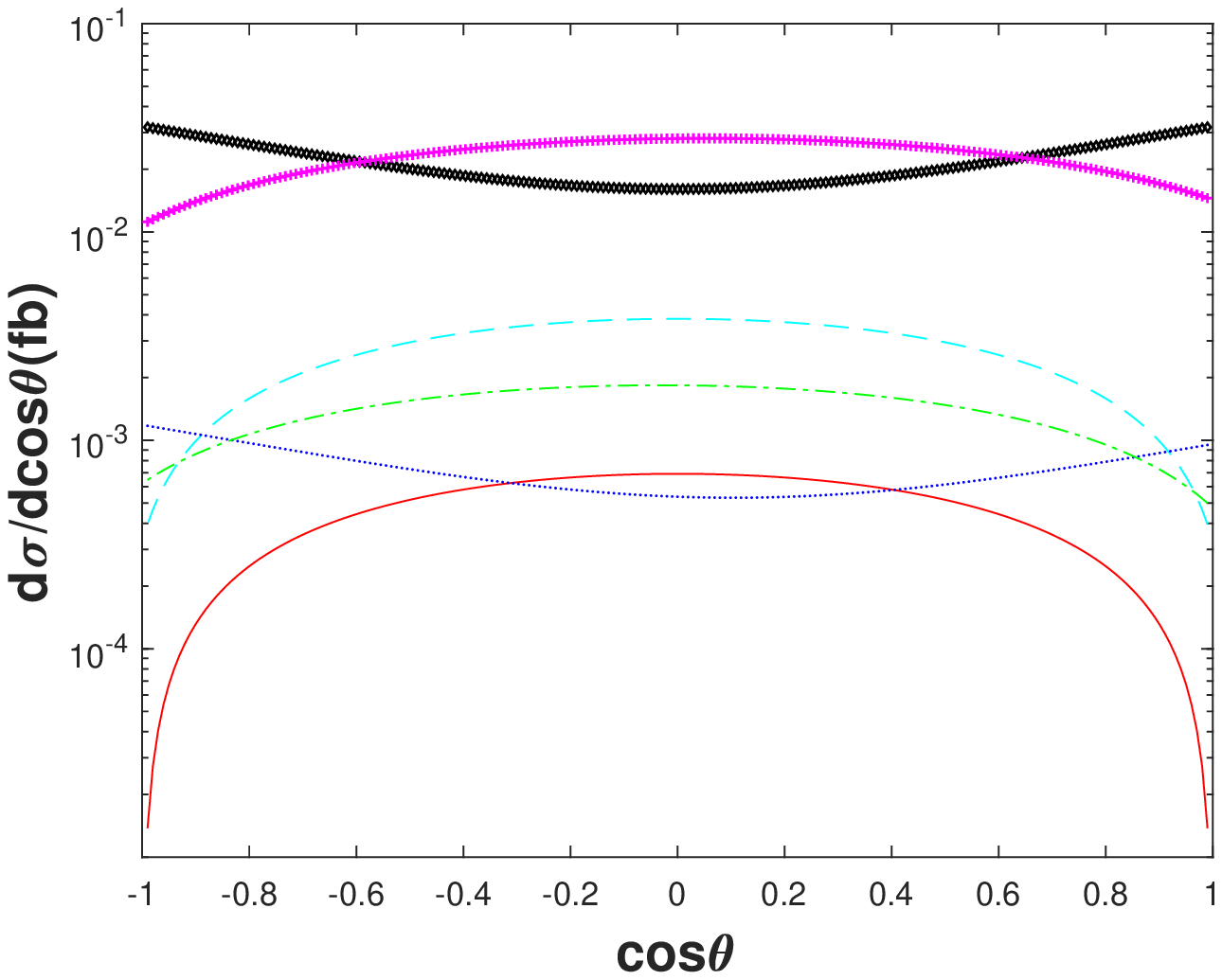}
\caption{
Differential cross sections d$\sigma$/dcos$\theta$ for the channel $e^-e^+\to \gamma^*/Z^0 \to|(b\bar{b})[n]\rangle+|(b\bar{b})[n']\rangle$ via the virtual photon $\gamma^*$ (left), the $Z^0$ boson (middle) and (right). The diamond black line, cross magenta line, dashed cyan line, solid red line, dotted blue line, and the dash-dotted green line on the left are for the double heavy quarkonium $|(b\bar{b})[^1S_0]\rangle/|(b\bar{b})[^3S_1]\rangle$, $|(b\bar{b})[^3S_1]\rangle/|(b\bar{b})[^1P_1]\rangle$,
$|(b\bar{b})[^1S_0]\rangle/|(b\bar{b})[^1P_1]\rangle$, $|(b\bar{b})[^3S_1]\rangle/|(b\bar{b})[^3P_0]\rangle$,
$|(b\bar{b})[^3S_1]\rangle/|(b\bar{b})[^3P_1]\rangle$, and $|(b\bar{b})[^3S_1]\rangle/|(b\bar{b})[^3P_2]\rangle$, respectively.
That on the middle are for the double heavy quarkonium $|(b\bar{b})[^1S_0]\rangle/|(b\bar{b})[^3S_1]\rangle$, $|(b\bar{b})[^3S_1]\rangle/|(b\bar{b})[^3S_1]\rangle$, $|(b\bar{b})[^3S_1]\rangle/|(b\bar{b})[^1P_1]\rangle$, $|(b\bar{b})[^3S_1]\rangle/|(b\bar{b})[^3P_0]\rangle$,
$|(b\bar{b})[^3S_1]\rangle/|(b\bar{b})[^3P_1]\rangle$, and $|(b\bar{b})[^3S_1]\rangle/|(b\bar{b})[^3P_2]\rangle$, respectively.
That on the right are for the double heavy quarkonium $|(b\bar{b})[^1S_0]\rangle/|(b\bar{b})[^3S_1]\rangle$, $|(b\bar{b})[^3S_1]\rangle/|(b\bar{b})[^3S_1]\rangle$, $|(b\bar{b})[^1S_0]\rangle/|(b\bar{b})[^1P_1]\rangle$, $|(b\bar{b})[^1S_0]\rangle/|(b\bar{b})[^3P_0]\rangle$,
$|(b\bar{b})[^1S_0]\rangle/|(b\bar{b})[^3P_1]\rangle$, and $|(b\bar{b})[^1S_0]\rangle/|(b\bar{b})[^3P_2]\rangle$, respectively.
} \label{bbdcos}
\end{figure*}

\begin{figure*}[htbp]
\centering
\includegraphics[width=0.32\textwidth]{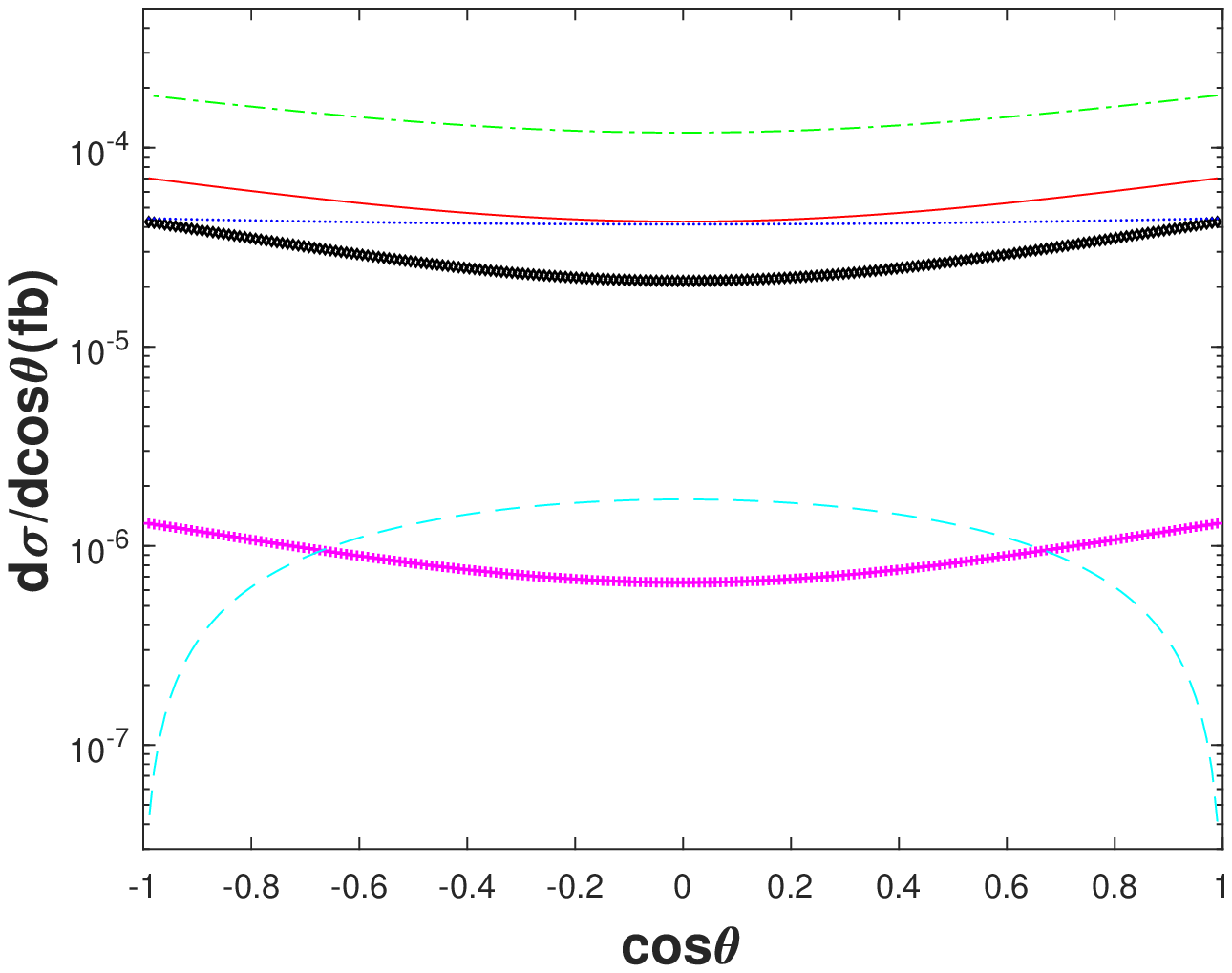}
\includegraphics[width=0.32\textwidth]{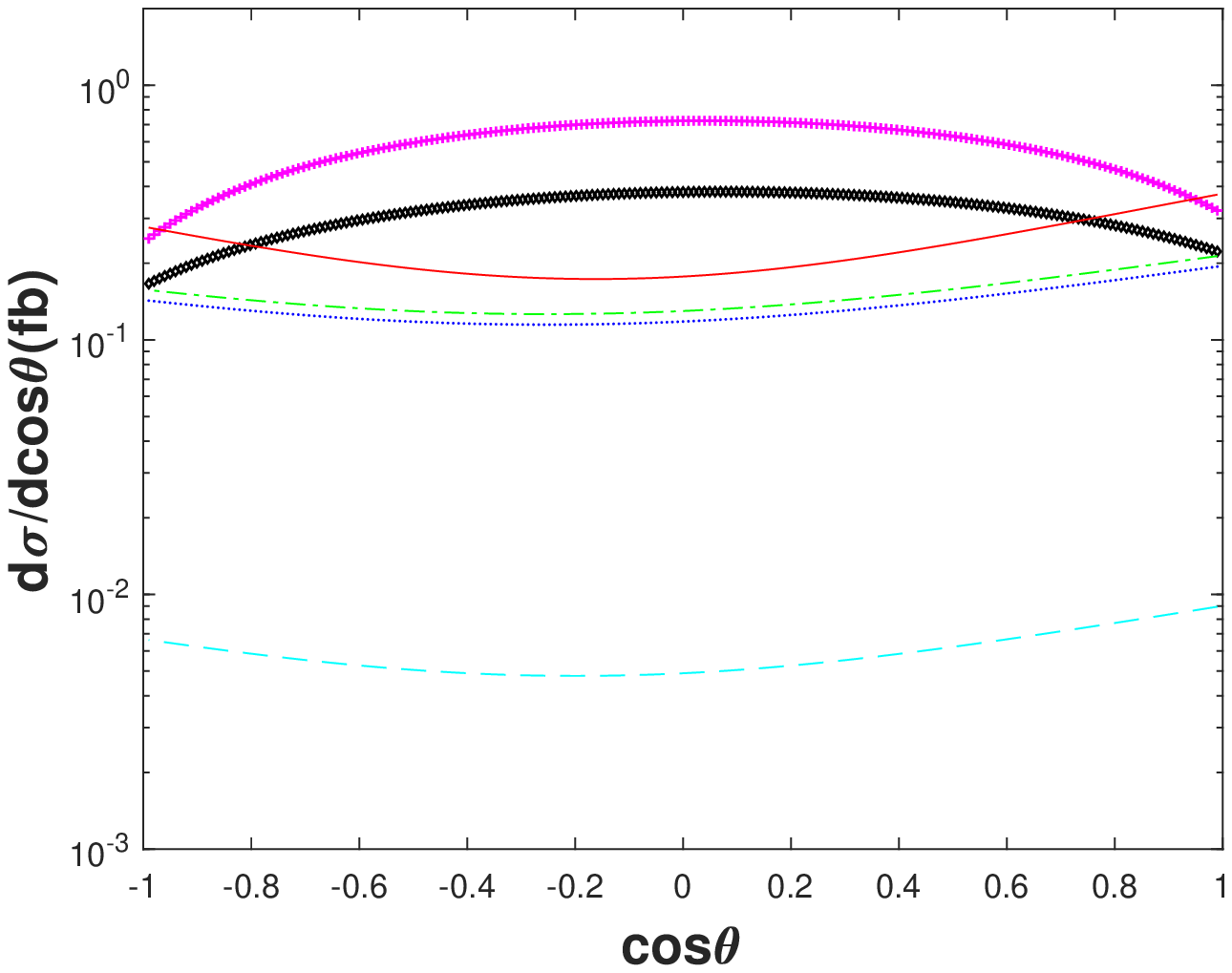}
\includegraphics[width=0.32\textwidth]{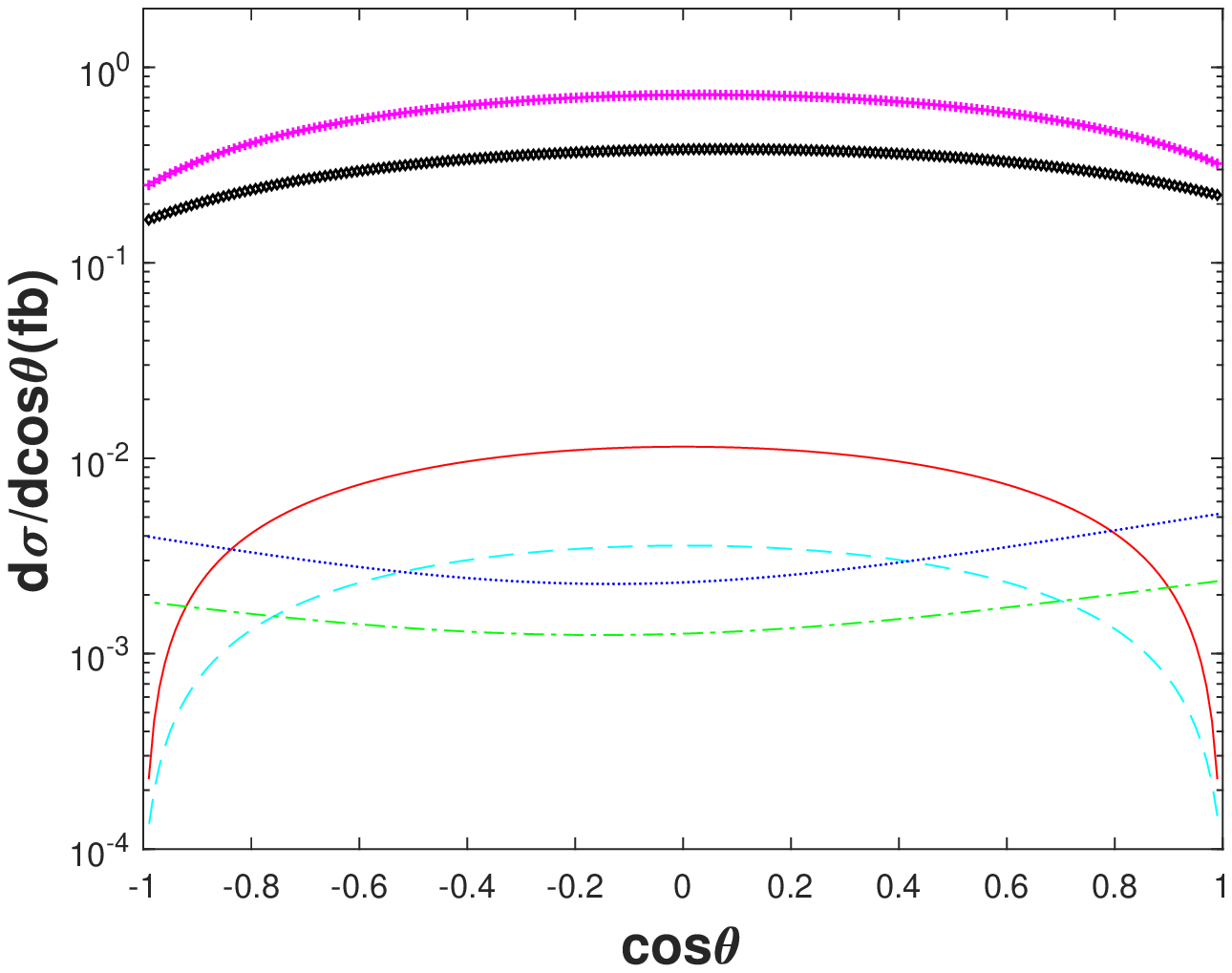}
\caption{
Differential cross sections d$\sigma$/dcos$\theta$ for the channel $e^-e^+\to \gamma^*/Z^0 \to|(c\bar{b})[n]\rangle+|(b\bar{c})[n']\rangle$ via the virtual photon $\gamma^*$ (left), the $Z^0$ boson (middle) and (right). The diamond black line, cross magenta line, dashed cyan line, solid red line, dotted blue line, and the dash-dotted green line on the left are for the double heavy quarkonium $|(c\bar{b})[^1S_0]\rangle/|(b\bar{c})[^3S_1]\rangle$, $|(c\bar{b})[^3S_1]\rangle/|(b\bar{c})[^1P_1]\rangle$,
$|(c\bar{b})[^1S_0]\rangle/|(b\bar{c})[^1P_1]\rangle$, $|(c\bar{b})[^3S_1]\rangle/|(b\bar{c})[^3P_0]\rangle$,
$|(c\bar{b})[^3S_1]\rangle/|(b\bar{c})[^3P_1]\rangle$, and $|(c\bar{b})[^3S_1]\rangle/|(b\bar{c})[^3P_2]\rangle$, respectively.
That on the middle are for the double heavy quarkonium $|(c\bar{b})[^1S_0]\rangle/|(b\bar{c})[^3S_1]\rangle$, $|(c\bar{b})[^3S_1]\rangle/|(b\bar{c})[^3S_1]\rangle$, $|(c\bar{b})[^3S_1]\rangle/|(b\bar{c})[^1P_1]\rangle$, $|(c\bar{b})[^3S_1]\rangle/|(b\bar{c})[^3P_0]\rangle$,
$|(c\bar{b})[^3S_1]\rangle/|(b\bar{c})[^3P_1]\rangle$, and $|(c\bar{b})[^3S_1]\rangle/|(b\bar{c})[^3P_2]\rangle$, respectively.
That on the right are for the double heavy quarkonium $|(c\bar{b})[^1S_0]\rangle/|(b\bar{c})[^3S_1]\rangle$, $|(c\bar{b})[^3S_1]\rangle/|(b\bar{c})[^3S_1]\rangle$, $|(c\bar{b})[^1S_0]\rangle/|(b\bar{c})[^1P_1]\rangle$, $|(c\bar{b})[^1S_0]\rangle/|(b\bar{c})[^3P_0]\rangle$,
$|(c\bar{b})[^1S_0]\rangle/|(b\bar{c})[^3P_1]\rangle$, and $|(c\bar{b})[^1S_0]\rangle/|(b\bar{c})[^3P_2]\rangle$, respectively.
} \label{bcdcos}
\end{figure*}

In Figs. \ref{ccdcos}-\ref{bcdcos}, differential distributions $d\sigma/dcos\theta$ for double quarkonium $|(c\bar{c})[n]\rangle$~/$|(c\bar{c})[n']\rangle$, $|(b\bar{b})[n]\rangle$~/$|(b\bar{b})[n']\rangle$~, and $|(c\bar{b})[n]\rangle~/|(b\bar{c})[n']\rangle$ are displayed, where $[n]$ ~/$[n']$ stands for $^1S_0$,  $^3S_1$,  $^1P_1$, and $^3P_J$ ($J=0,1,2$).
Here, $\theta$ is the angle between the momentum $\vec{p_1}$ of electron and the momentum $\vec{q_1}$ of the heavy quarkonium.
It is shown that the $Z^0$ propagated processes and the corresponding virtual photon propagated ones have similar line shapes.
We also find that $d\sigma/dcos\theta$ approaches its maximum when the heavy quarkonium and the electron running in the same direction or back-to-back for both $S$-wave and $P$-wave states.

\begin{figure*}[htbp]
\centering
\includegraphics[width=0.32\textwidth]{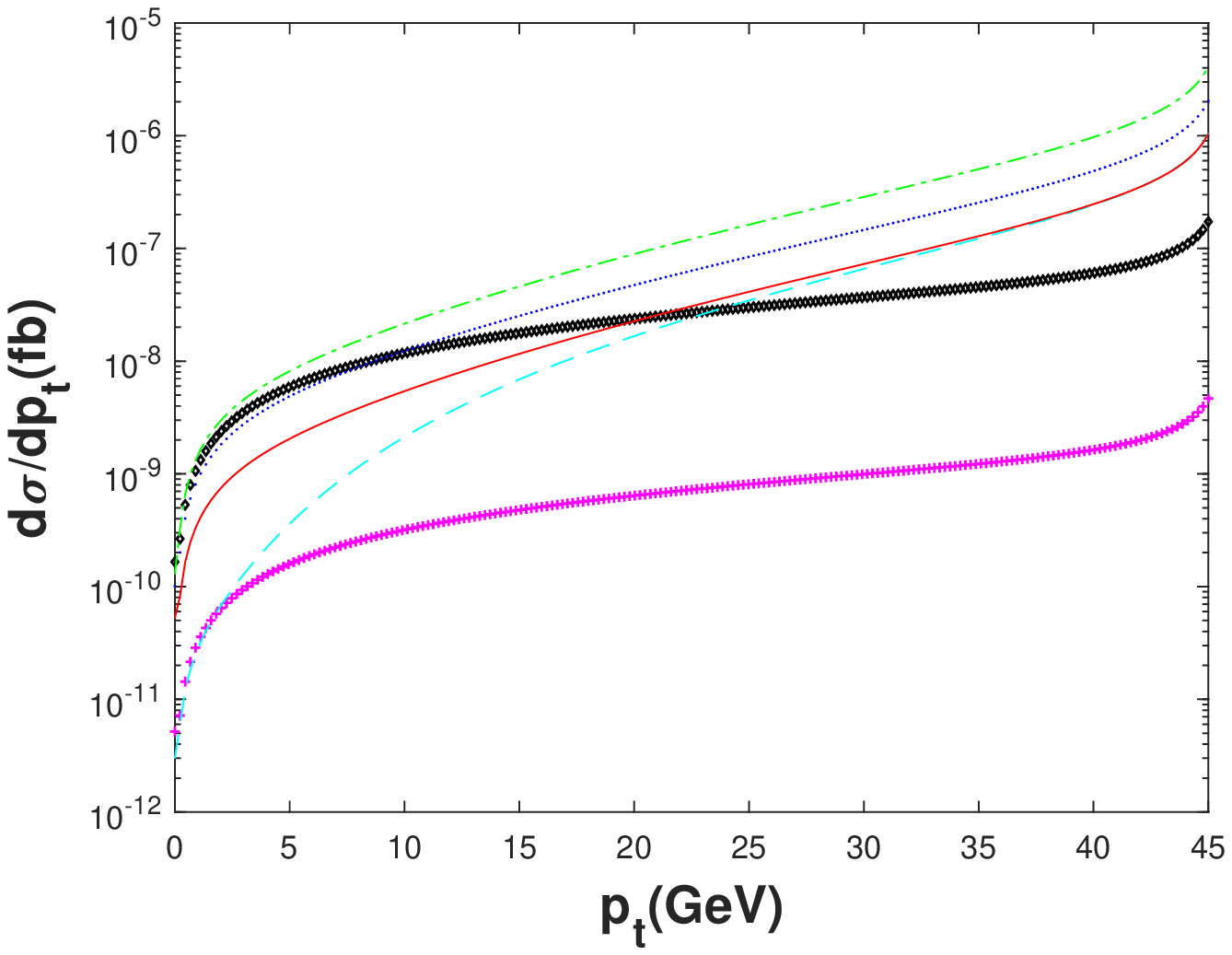}
\includegraphics[width=0.32\textwidth]{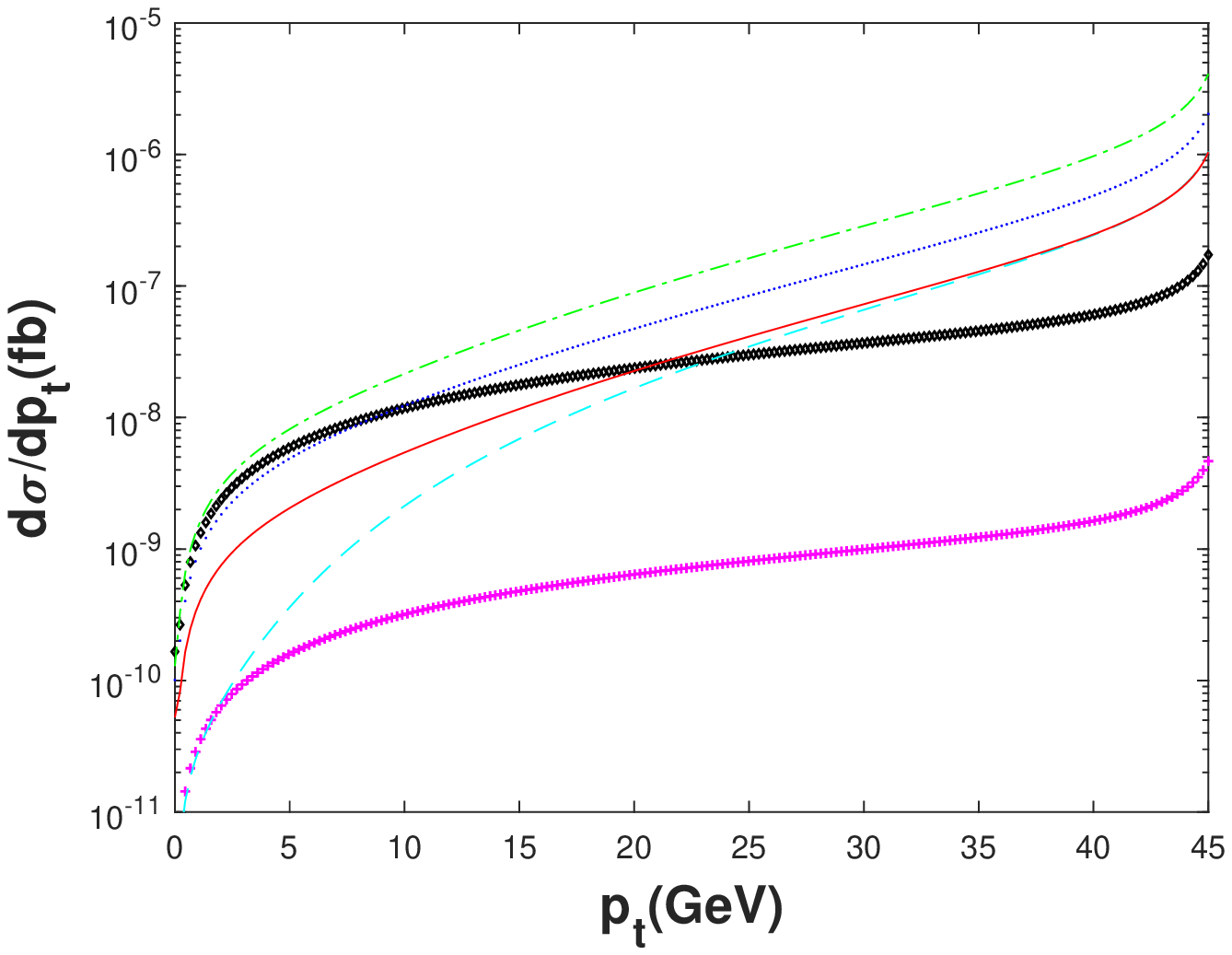}
\includegraphics[width=0.32\textwidth]{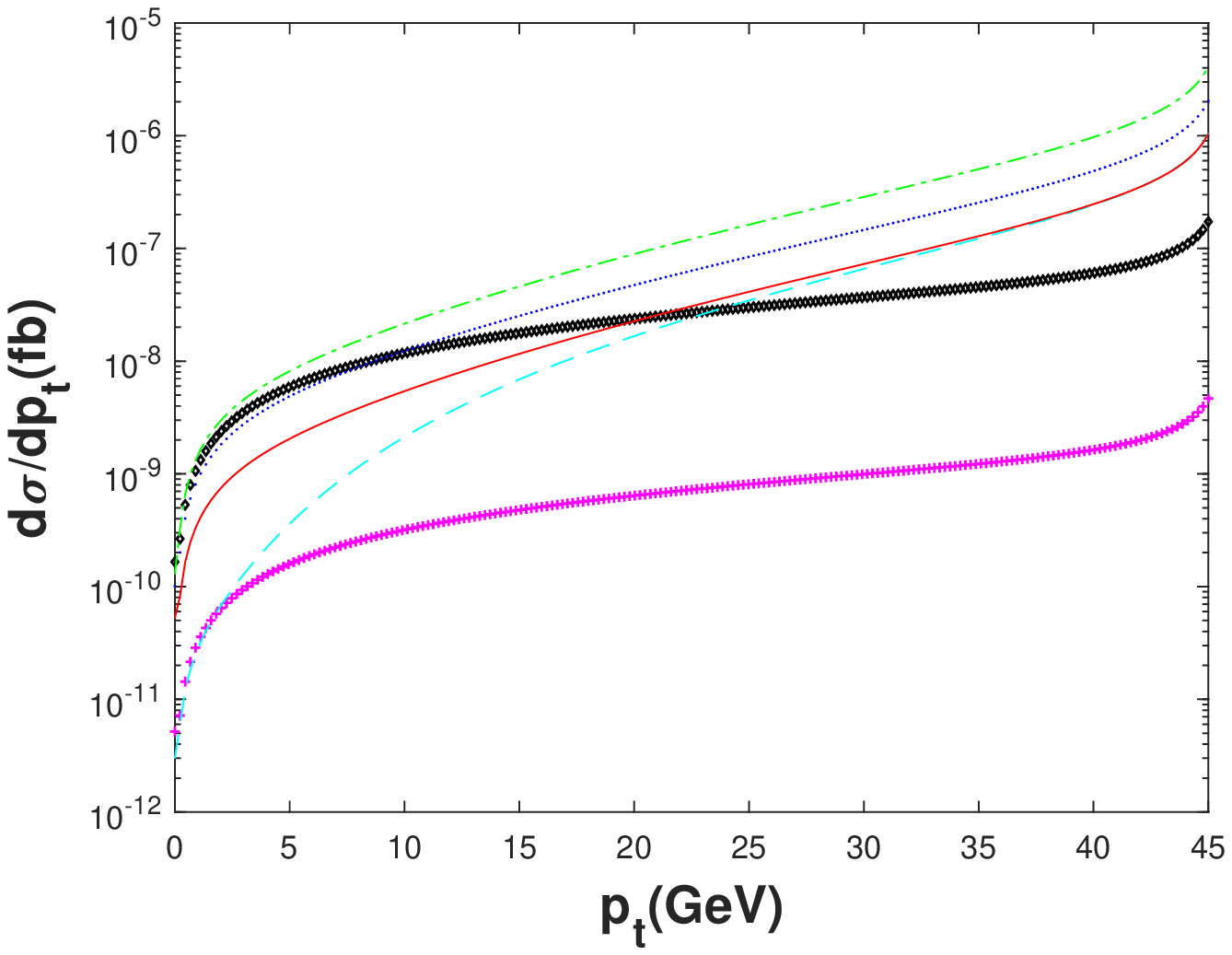}
\caption{
Differential cross sections d$\sigma$/d$p_t$ for the channel $e^-e^+\to \gamma^*/Z^0 \to|(c\bar{c})[n]\rangle+|(c\bar{c})[n']\rangle$ via the virtual photon $\gamma^*$ (left), the $Z^0$ boson (middle) and (right). The diamond black line, cross magenta line, dashed cyan line, solid red line, dotted blue line, and the dash-dotted green line on the left are for the double heavy quarkonium $|(c\bar{c})[^1S_0]\rangle/|(c\bar{c})[^3S_1]\rangle$, $|(c\bar{c})[^3S_1]\rangle/|(c\bar{c})[^1P_1]\rangle$,
$|(c\bar{c})[^1S_0]\rangle/|(c\bar{c})[^1P_1]\rangle$, $|(c\bar{c})[^3S_1]\rangle/|(c\bar{c})[^3P_0]\rangle$,
$|(c\bar{c})[^3S_1]\rangle/|(c\bar{c})[^3P_1]\rangle$, and $|(c\bar{c})[^3S_1]\rangle/|(c\bar{c})[^3P_2]\rangle$, respectively.
That on the middle are for the double heavy quarkonium $|(c\bar{c})[^1S_0]\rangle/|(c\bar{c})[^3S_1]\rangle$, $|(c\bar{c})[^3S_1]\rangle/|(c\bar{c})[^3S_1]\rangle$, $|(c\bar{c})[^3S_1]\rangle/|(c\bar{c})[^1P_1]\rangle$, $|(c\bar{c})[^3S_1]\rangle/|(c\bar{c})[^3P_0]\rangle$,
$|(c\bar{c})[^3S_1]\rangle/|(c\bar{c})[^3P_1]\rangle$, and $|(c\bar{c})[^3S_1]\rangle/|(c\bar{c})[^3P_2]\rangle$, respectively.
That on the right are for the double heavy quarkonium $|(c\bar{c})[^1S_0]\rangle/|(c\bar{c})[^3S_1]\rangle$, $|(c\bar{c})[^3S_1]\rangle/|(c\bar{c})[^3S_1]\rangle$, $|(c\bar{c})[^1S_0]\rangle/|(c\bar{c})[^1P_1]\rangle$, $|(c\bar{c})[^1S_0]\rangle/|(c\bar{c})[^3P_0]\rangle$,
$|(c\bar{c})[^1S_0]\rangle/|(c\bar{c})[^3P_1]\rangle$, and $|(c\bar{c})[^1S_0]\rangle/|(c\bar{c})[^3P_2]\rangle$, respectively.
} \label{ccdpt}
\end{figure*}

\begin{figure*}[htbp]
\centering
\includegraphics[width=0.32\textwidth]{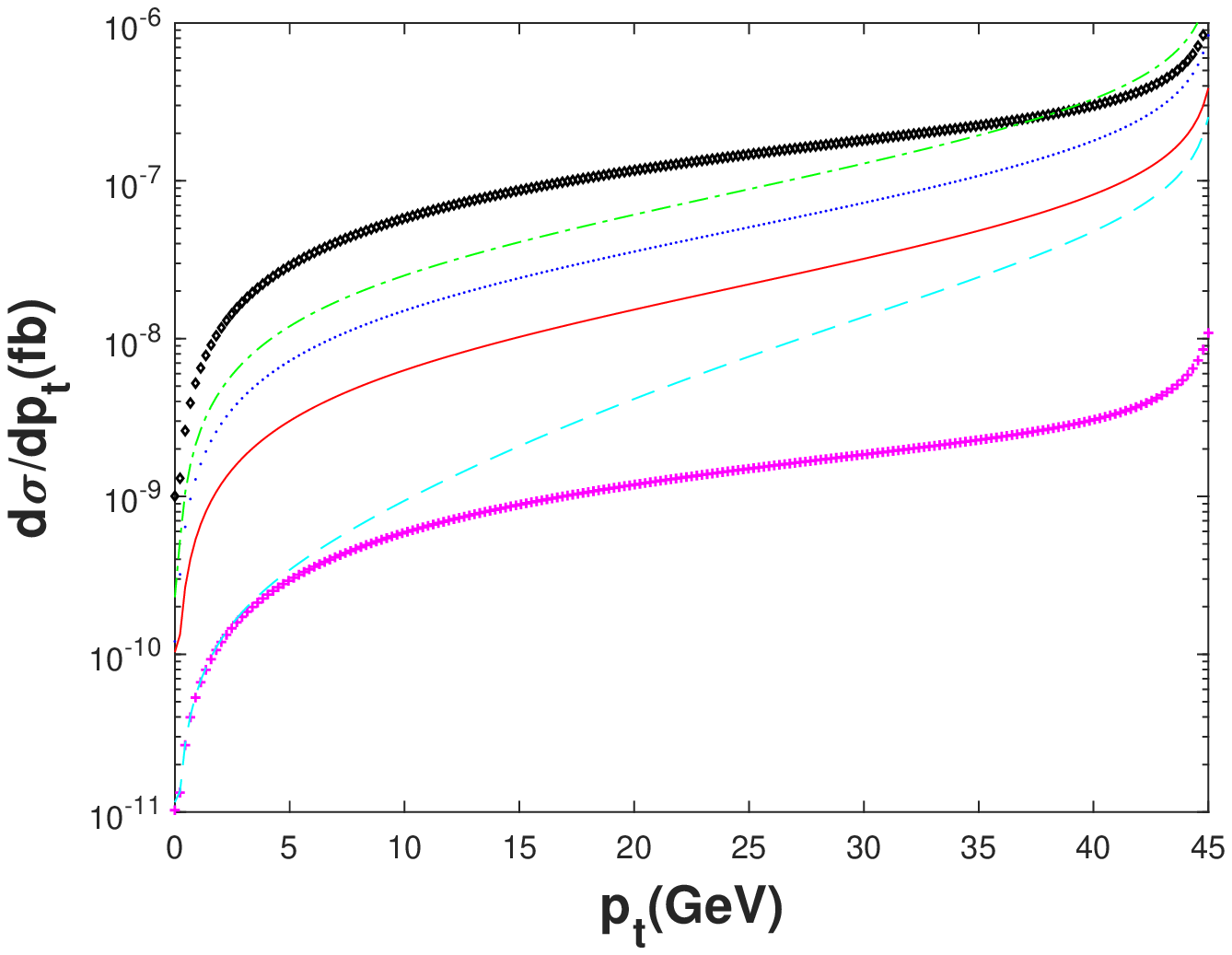}
\includegraphics[width=0.32\textwidth]{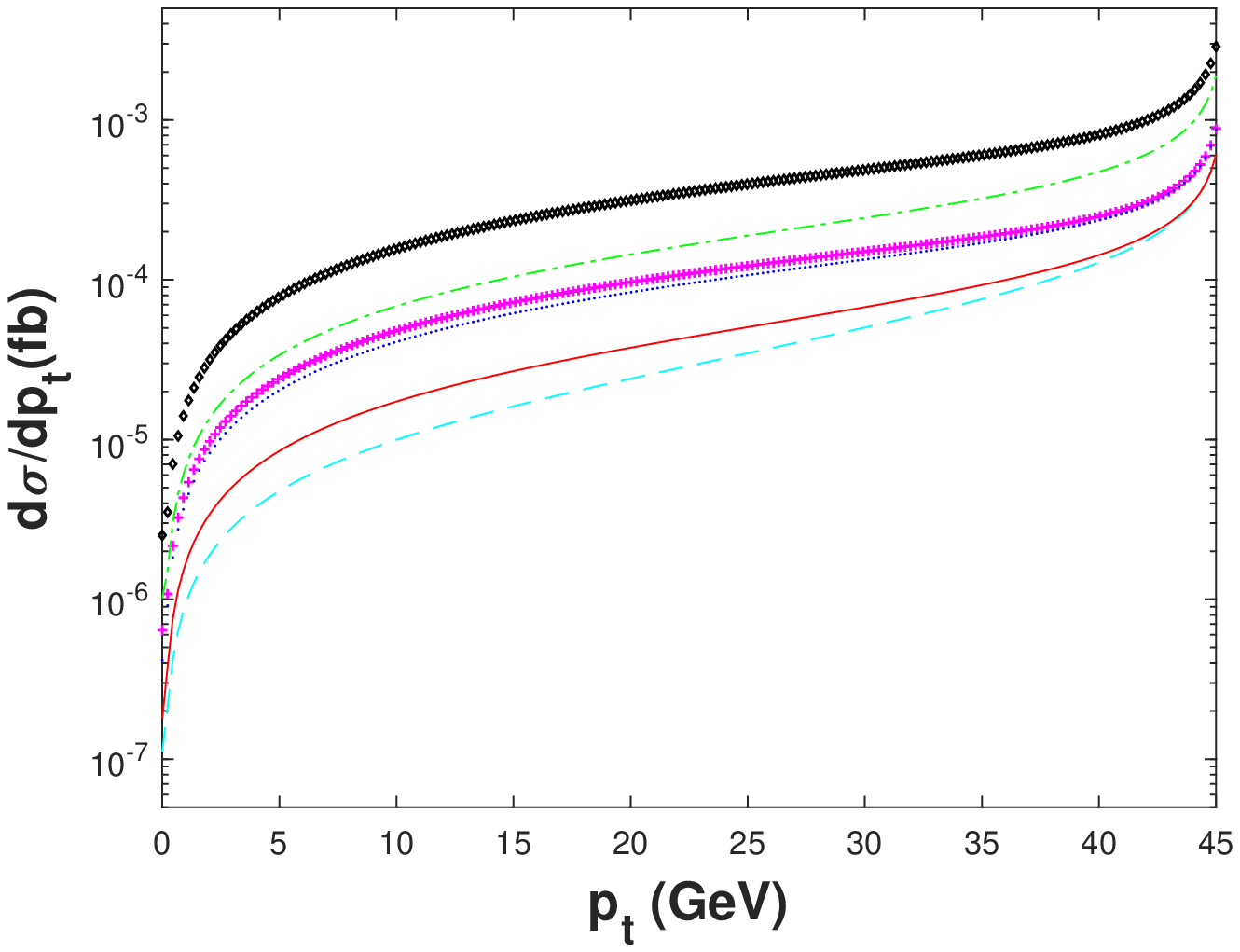}
\includegraphics[width=0.32\textwidth]{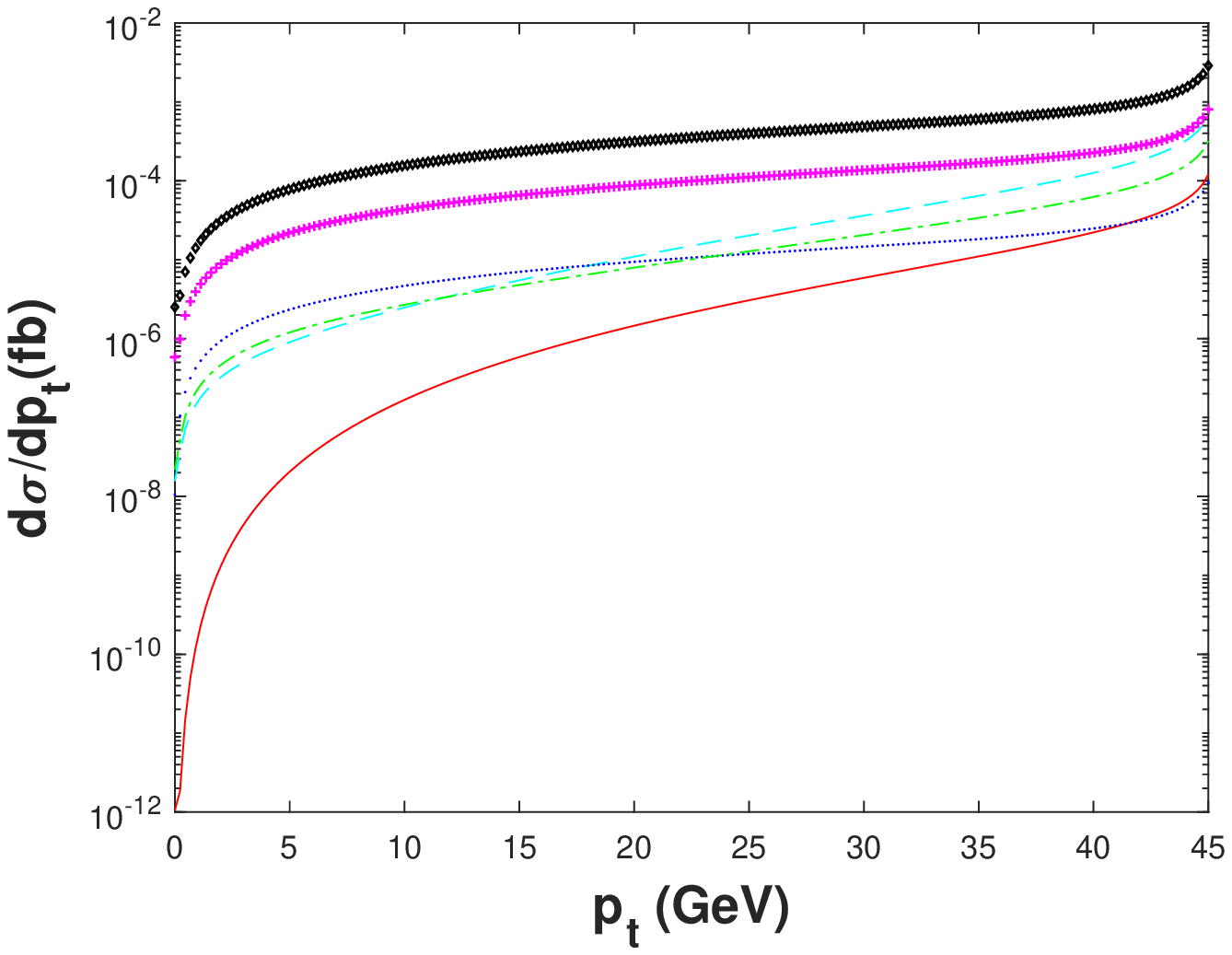}
\caption{
Differential cross sections d$\sigma$/d$p_t$ for the channel $e^-e^+\to \gamma^*/Z^0 \to|(b\bar{b})[n]\rangle+|(b\bar{b})[n']\rangle$ via the virtual photon $\gamma^*$ (left), the $Z^0$ boson (middle) and (right). The diamond black line, cross magenta line, dashed cyan line, solid red line, dotted blue line, and the dash-dotted green line on the left are for the double heavy quarkonium $|(b\bar{b})[^1S_0]\rangle/|(b\bar{b})[^3S_1]\rangle$, $|(b\bar{b})[^3S_1]\rangle/|(b\bar{b})[^1P_1]\rangle$,
$|(b\bar{b})[^1S_0]\rangle/|(b\bar{b})[^1P_1]\rangle$, $|(b\bar{b})[^3S_1]\rangle/|(b\bar{b})[^3P_0]\rangle$,
$|(b\bar{b})[^3S_1]\rangle/|(b\bar{b})[^3P_1]\rangle$, and $|(b\bar{b})[^3S_1]\rangle/|(b\bar{b})[^3P_2]\rangle$, respectively.
That on the middle are for the double heavy quarkonium $|(b\bar{b})[^1S_0]\rangle/|(b\bar{b})[^3S_1]\rangle$, $|(b\bar{b})[^3S_1]\rangle/|(b\bar{b})[^3S_1]\rangle$, $|(b\bar{b})[^3S_1]\rangle/|(b\bar{b})[^1P_1]\rangle$, $|(b\bar{b})[^3S_1]\rangle/|(b\bar{b})[^3P_0]\rangle$,
$|(b\bar{b})[^3S_1]\rangle/|(b\bar{b})[^3P_1]\rangle$, and $|(b\bar{b})[^3S_1]\rangle/|(b\bar{b})[^3P_2]\rangle$, respectively.
That on the right are for the double heavy quarkonium $|(b\bar{b})[^1S_0]\rangle/|(b\bar{b})[^3S_1]\rangle$, $|(b\bar{b})[^3S_1]\rangle/|(b\bar{b})[^3S_1]\rangle$, $|(b\bar{b})[^1S_0]\rangle/|(b\bar{b})[^1P_1]\rangle$, $|(b\bar{b})[^1S_0]\rangle/|(b\bar{b})[^3P_0]\rangle$,
$|(b\bar{b})[^1S_0]\rangle/|(b\bar{b})[^3P_1]\rangle$, and $|(b\bar{b})[^1S_0]\rangle/|(b\bar{b})[^3P_2]\rangle$, respectively.
} \label{bbdpt}
\end{figure*}

\begin{figure*}[htbp]
\centering
\includegraphics[width=0.32\textwidth]{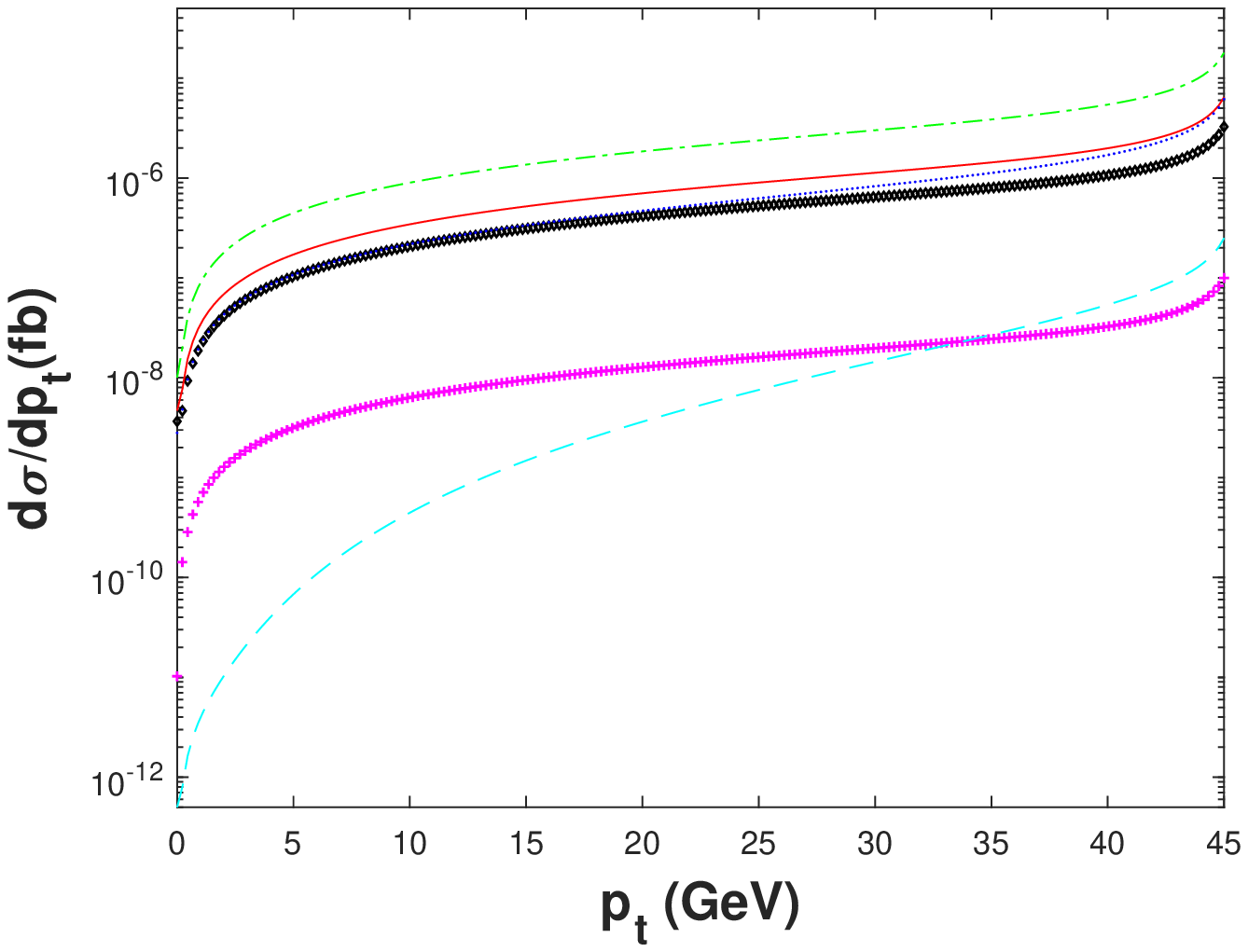}
\includegraphics[width=0.32\textwidth]{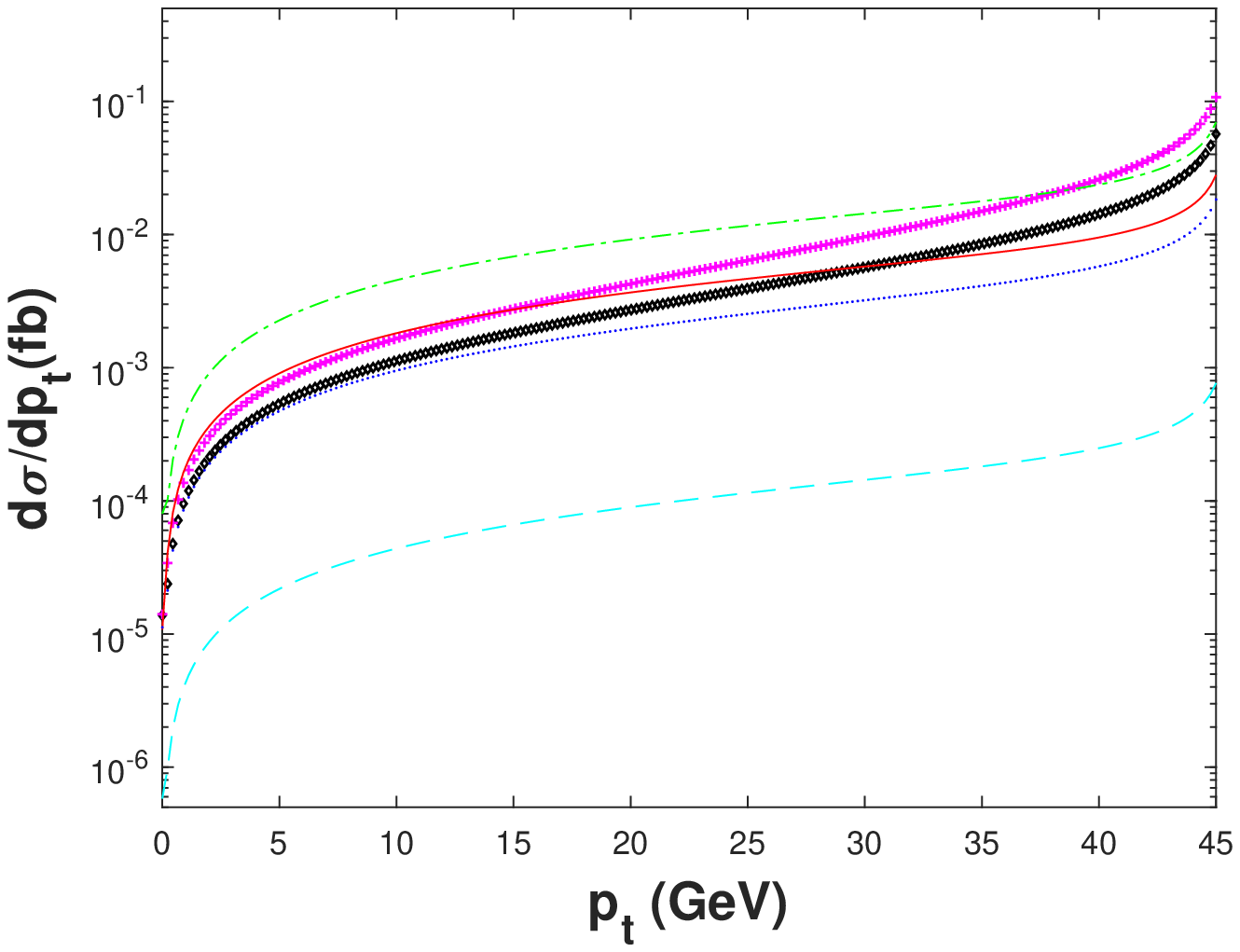}
\includegraphics[width=0.32\textwidth]{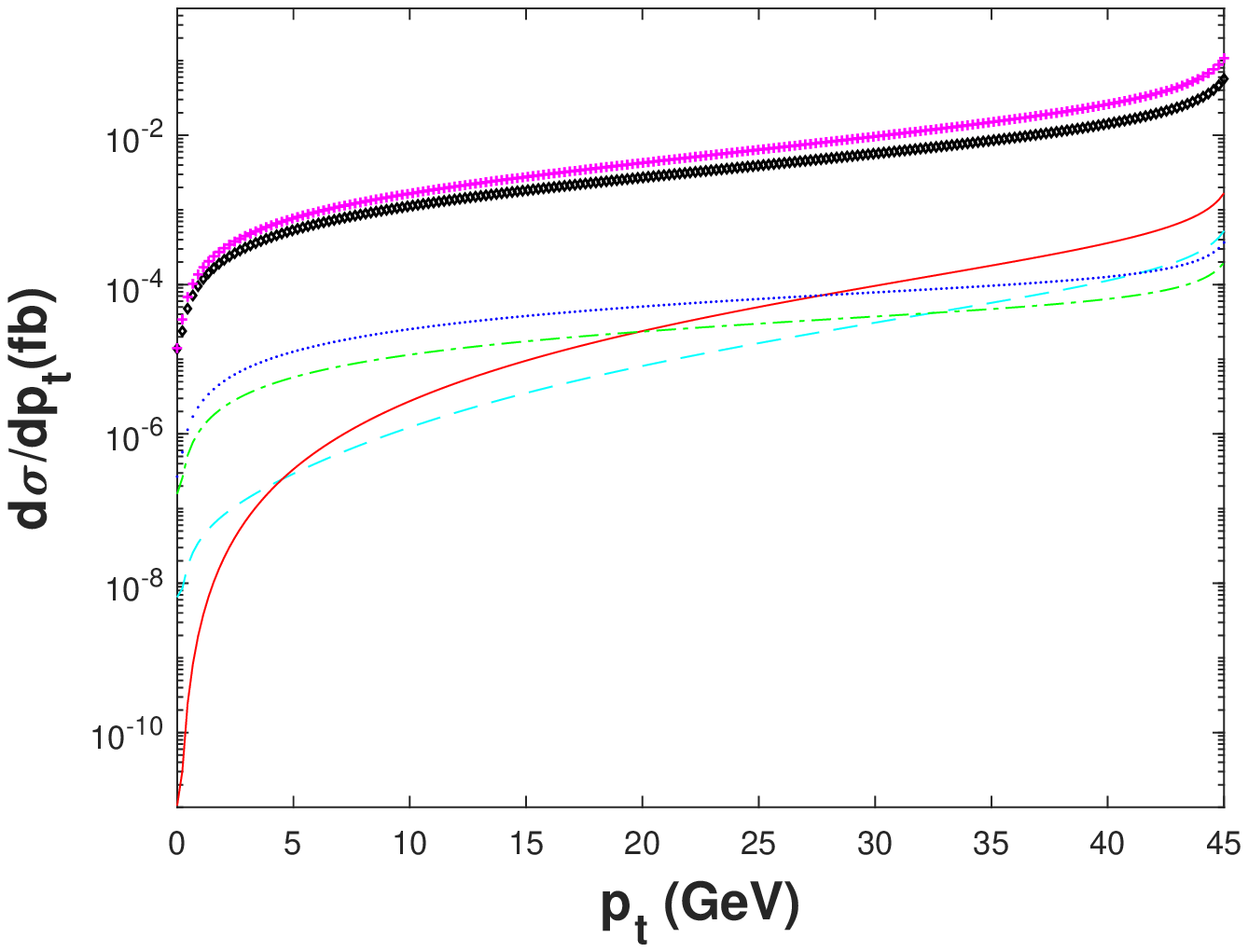}
\caption{
Differential cross sections d$\sigma$/d$p_t$ for the channel $e^-e^+\to \gamma^*/Z^0 \to|(c\bar{b})[n]\rangle+|(b\bar{c})[n']\rangle$ via the virtual photon $\gamma^*$ (left), the $Z^0$ boson (middle) and (right). The diamond black line, cross magenta line, dashed cyan line, solid red line, dotted blue line, and the dash-dotted green line on the left are for the double heavy quarkonium $|(c\bar{b})[^1S_0]\rangle/|(b\bar{c})[^3S_1]\rangle$, $|(c\bar{b})[^3S_1]\rangle/|(b\bar{c})[^1P_1]\rangle$,
$|(c\bar{b})[^1S_0]\rangle/|(b\bar{c})[^1P_1]\rangle$, $|(c\bar{b})[^3S_1]\rangle/|(b\bar{c})[^3P_0]\rangle$,
$|(c\bar{b})[^3S_1]\rangle/|(b\bar{c})[^3P_1]\rangle$, and $|(c\bar{b})[^3S_1]\rangle/|(b\bar{c})[^3P_2]\rangle$, respectively.
That on the middle are for the double heavy quarkonium $|(c\bar{b})[^1S_0]\rangle/|(b\bar{c})[^3S_1]\rangle$, $|(c\bar{b})[^3S_1]\rangle/|(b\bar{c})[^3S_1]\rangle$, $|(c\bar{b})[^3S_1]\rangle/|(b\bar{c})[^1P_1]\rangle$, $|(c\bar{b})[^3S_1]\rangle/|(b\bar{c})[^3P_0]\rangle$,
$|(c\bar{b})[^3S_1]\rangle/|(b\bar{c})[^3P_1]\rangle$, and $|(c\bar{b})[^3S_1]\rangle/|(b\bar{c})[^3P_2]\rangle$, respectively.
That on the right are for the double heavy quarkonium $|(c\bar{b})[^1S_0]\rangle/|(b\bar{c})[^3S_1]\rangle$, $|(c\bar{b})[^3S_1]\rangle/|(b\bar{c})[^3S_1]\rangle$, $|(c\bar{b})[^1S_0]\rangle/|(b\bar{c})[^1P_1]\rangle$, $|(c\bar{b})[^1S_0]\rangle/|(b\bar{c})[^3P_0]\rangle$,
$|(c\bar{b})[^1S_0]\rangle/|(b\bar{c})[^3P_1]\rangle$, and $|(c\bar{b})[^1S_0]\rangle/|(b\bar{c})[^3P_2]\rangle$, respectively.
} \label{bcdpt}
\end{figure*}

The transverse momentum $p_{t}$ distribution of the heavy quarkonium can further tell us more information on the production of double charmonium, double bottomonium, double $Bc$-mesons. If the distribution $d\sigma/dcos\theta$ is set to be
\begin{eqnarray}
\frac{d\sigma}{dcos\theta}=f(cos\theta),
\end{eqnarray}
Which can be easily obtained with the differential phase space of Eq. (\ref{dPhi-2}), then the distribution $d\sigma/dp_t$ can be
\begin{eqnarray}
\frac{d\sigma}{dp_t}&=&\left|\frac{dcos\theta}{dp_t}\right| \left(\frac{d\sigma}{dcos\theta}\right)\nonumber\\
&=&\frac{p_t}{|\vec{q}_1| \sqrt{|\vec{q}_1|^2-p^2_t}} f(cos\theta),
\end{eqnarray}
Where ${|\vec{q}_1|}=\sqrt{\lambda[s, M^2_{Q\bar{Q'}}, M^2_{Q'\bar{Q}}]} /2\sqrt{s}$ is the magnitude of the momentum of the heavy quarkonium.
We present the transverse momentum $p_t$ distributions for the cross sections in Fig. \ref{ccdpt}-\ref{bcdpt} for double states $|(c\bar{c})[1]\rangle$, $|(b\bar{b})[1]\rangle$, and $|(c\bar{b})[1]\rangle/|(b\bar{c})[1]\rangle$.
Since the differential distribution is proportional to $p_t/\sqrt{|\vec{q}_1|^2-p^2_t}$ and values of the function $f(cos\theta)$ changes smoothly, $d\sigma/dp_t$ shall increase with the increment of transverse momentum $p_t$.

\subsection{Uncertainty analysis}
\label{uncertainty}
\begin{table}
\caption{Cross sections (units:~$\times 10^{-5}~fb$) for the production of double quarkonium $|(c\bar{c})[n]\rangle$ in $e^+e^-$ annihilation via $Z^0$ at the center-of-mass energy $\sqrt{s}=91.1876$GeV within the BT-potential model ($n_f=3$)~\cite{lx}.}
\begin{tabular}{|c|c|}
\hline\hline
$\sigma{(|(c\bar{c})[^1S_0]\rangle+|(c\bar{c})[^3S_1]\rangle)}$&~$18.95^{+3.64}_{-4.70}$~\\
\hline
$\sigma{(|(c\bar{c})[^3S_1]\rangle+|(c\bar{c})[^3S_1]\rangle)}$&~$64.16^{+12.29}_{-15.87}$~\\
\hline
$\sigma{(|(c\bar{c})[^1S_0]\rangle+|(c\bar{c})[^1P_1]\rangle)}$&~$60.76^{+4.49}_{-7.35}$~\\
\hline
$\sigma{(|(c\bar{c})[^3S_1]\rangle+|(c\bar{c})[^1P_1]\rangle)}$&~$454.0^{+39.1}_{-59.2}$~\\
\hline
$\sigma{(|(c\bar{c})[^1S_0]\rangle+|(c\bar{c})[^3P_0]\rangle)}$&~$137.0^{+9.9}_{-16.4}$~\\
\hline
$\sigma{(|(c\bar{c})[^1S_0]\rangle+|(c\bar{c})[^3P_1]\rangle)}$&~$18.87^{+3.98}_{-4.27}$~\\
\hline
$\sigma{(|(c\bar{c})[^1S_0]\rangle+|(c\bar{c})[^3P_2]\rangle)}$&~$284.2^{+22.2}_{-35.3}$~\\
\hline
$\sigma{(|(c\bar{c})[^3S_1]\rangle+|(c\bar{c})[^3P_0]\rangle)}$&~$44.03^{+4.60}_{-6.38}$~\\
\hline
$\sigma{(|(c\bar{c})[^3S_1]\rangle+|(c\bar{c})[^3P_1]\rangle)}$&~$17.62^{+3.57}_{-3.88}$~\\
\hline
$\sigma{(|(c\bar{c})[^3S_1]\rangle+|(c\bar{c})[^3P_2]\rangle)}$&~$57.51^{+8.25}_{-10.02}$~\\
\hline
Sum&~$1158^{+112}_{-164}$~\\
\hline\hline
\end{tabular}
\label{tabrpd}
\end{table}

\begin{table}
\caption{Cross sections (units:~$\times 10^{-4}~fb$) for the production of double quarkonium $|(b\bar{b})[n]\rangle$ in $e^+e^-$ annihilation via $Z^0$ at the center-of-mass energy $\sqrt{s}=91.187$6 $GeV$ within the BT-potential model ($n_f=4$)~\cite{lx}.}
\begin{tabular}{|c|c|}
\hline\hline
$\sigma{(|(b\bar{b})[^1S_0]\rangle+|(b\bar{b})[^3S_1]\rangle)}$&~$428.6^{+67.9}_{-60.9}$~\\
\hline
$\sigma{(|(b\bar{b})[^3S_1]\rangle+|(b\bar{b})[^3S_1]\rangle)}$&~$119.9^{+18.4}_{-16.6}$~\\
\hline
$\sigma{(|(b\bar{b})[^1S_0]\rangle+|(b\bar{b})[^1P_1]\rangle)}$&~$53.27^{+1.46}_{-1.51}$~\\
\hline
$\sigma{(|(b\bar{b})[^3S_1]\rangle+|(b\bar{b})[^1P_1]\rangle)}$&~$61.46^{+4.22}_{-4.05}$~\\
\hline
$\sigma{(|(b\bar{b})[^1S_0]\rangle+|(b\bar{b})[^3P_0]\rangle)}$&~$9.215^{+0.141}_{-0.160}$~\\
\hline
$\sigma{(|(b\bar{b})[^1S_0]\rangle+|(b\bar{b})[^3P_1]\rangle)}$&~$14.31^{+1.60}_{-1.60}$~\\
\hline
$\sigma{(|(b\bar{b})[^1S_0]\rangle+|(b\bar{b})[^3P_2]\rangle)}$&~$28.15^{+1.38}_{-1.35}$~\\
\hline
$\sigma{(|(b\bar{b})[^3S_1]\rangle+|(b\bar{b})[^3P_0]\rangle)}$&~$68.37^{+6.17}_{-5.83}$~\\
\hline
$\sigma{(|(b\bar{b})[^3S_1]\rangle+|(b\bar{b})[^3P_1]\rangle)}$&~$117.5^{+13.6}_{-15.0}$~\\
\hline
$\sigma{(|(b\bar{b})[^3S_1]\rangle+|(b\bar{b})[^3P_2]\rangle)}$&~$230.6^{+23.6}_{-22.2}$~\\
\hline
Sum&~$1132^{+139}_{-129}$~\\
\hline\hline
\end{tabular}
\label{tabrpbe}
\end{table}

\begin{table}
\caption{Cross sections (units:~$\times 10^{-3}~fb$) for the production of double quarkonium  $|(c\bar{b})[n]\rangle$~/$|(b\bar{c})[n']\rangle$ in $e^+e^-$ annihilation via $Z^0$  at the center-of-mass energy $\sqrt{s}=91.187$6 $GeV$ within the BT-potential model ($n_f=3$)~\cite{lx}.}
\begin{tabular}{|c|c|}
\hline\hline
$\sigma{(|(c\bar{b})[^1S_0]\rangle+|(b\bar{c})[^3S_1]\rangle)}$&~$634.6^{+47.8}_{-45.8}$~~\\
\hline
$\sigma{(|(c\bar{b})[^3S_1]\rangle+|(b\bar{c})[^3S_1]\rangle)}$&~$1150^{+87}_{-84}$~\\
\hline
$\sigma{(|(c\bar{b})[^1S_0]\rangle+|(b\bar{c})[^1P_1]\rangle)}$&~$4.810^{+0.399}_{-0.406}$~\\
\hline
$\sigma{(|(c\bar{b})[^3S_1]\rangle+|(b\bar{c})[^1P_1]\rangle)}$&~$11.77^{+1.10}_{-1.05}$~\\
\hline
$\sigma{(|(c\bar{b})[^1S_0]\rangle+|(b\bar{c})[^3P_0]\rangle)}$&~$15.27^{+1.10}_{-1.12}$~\\
\hline
$\sigma{(|(c\bar{b})[^1S_0]\rangle+|(b\bar{c})[^3P_1]\rangle)}$&~$6.162^{+0.701}_{-0.448}$~\\
\hline
$\sigma{(|(c\bar{b})[^1S_0]\rangle+|(b\bar{c})[^3P_2]\rangle)}$&~$3.093^{+0.253}_{-0.236}$~\\
\hline
$\sigma{(|(c\bar{b})[^3S_1]\rangle+|(b\bar{c})[^3P_0]\rangle)}$&~$454.2^{-15.7}_{+30.8}$~\\
\hline
$\sigma{(|(c\bar{b})[^3S_1]\rangle+|(b\bar{c})[^3P_1]\rangle)}$&~$270.1^{+16.2}_{-15.8}$~\\
\hline
$\sigma{(|(c\bar{b})[^3S_1]\rangle+|(b\bar{c})[^3P_2]\rangle)}$&~$1133^{-16}_{+22}$~\\
\hline
Sum&~$3692^{+124}_{-97}$~\\
\hline\hline
\end{tabular}
\label{tabrpf}
\end{table}
As the cross sections for the production of double quarkonium $|(Q\bar{Q'})[n]\rangle~/|(Q'\bar{Q})[n']\rangle$ in $e^+e^-$ annihilation via $\gamma^*$ at the center-of-mass energy $\sqrt{s}=91.187$6 $GeV$ is three to four orders of magnitude smaller than the cross sections in $e^+e^-$ annihilation via $Z^0$. We only discuss the uncertainty of cross sections for the production of double quarkonium$|(Q\bar{Q'})[n]\rangle~/|(Q'\bar{Q})[n']\rangle$ in $e^+e^-$ annihilation via $Z^0$.

For the leading-order calculation, the main uncertainty sources of cross sections include the fine structure constant $\alpha$, the running coupling constant $\alpha_s$, the Fermi constant $G_F$, the Weinberg angle $\theta_W$, the mass and width of  the $Z^0$ boson, the non-perturbative matrix elements, and the masses of constituent quarks.
Since parameters $\alpha$, $\alpha_s$, $G_F$, $\theta_W$,  and the mass and width of  the $Z^0$ boson are either an overall factor or an relatively precise value, we will not discuss uncertainties caused by them.
In this subsection, we will explore uncertainties caused by masses of constituent quarks, the non-perturbative matrix elements, and deviation of CM energy $\sqrt{s}$ away from $m_Z$.

The uncertainties of cross sections caused by varying the masses of constituent quarks by 0.1 GeV for $m_c$ and 0.2 GeV for $m_b$ (as shown in Table \ref{M&R} ) at the CM energy $\sqrt{s}=91.1876$ GeV are presented in Tables~~\ref{tabrpd}--\ref{tabrpf} for $Z^0$ propagated processes.
It worth noting that the effects of uncertainties of radial wave functions at the origin and their first derivatives at the origin caused by varying masses are also taken into consideration.
It is found that the wave functions at the origin and their derivatives at the origin increase as quark masses increase.
But, we find that the short-distance coefficients decrease along with the increasement of quark masses.
The overall effect is that the cross sections decrease with the increment of the quark masses.

The other four potential models is adopted to estimate the uncertainties caused by the wave functions at the origin and their first derivatives at the origin in Tables \ref{wavefunc1}- \ref{wavefunc3} for $Z^0$ propagated processes for double charmonium, double bottomonium, and double $Bc$ mesons, respectively.
The four models are QCD-motivated potential with one-loop correction given by John L. Richardson (J. potential) \cite{Richardson:1978bt},
QCD-motivated potential with two-loop correction given by K. Igi and S. Ono (I.O. potential) \cite{IO,Ikhdair:2003ry}, QCD-motivated potential with two-loop correction given by Yu-Qi Chen and Yu-Ping Kuang (C.K. potential) \cite{Chen:1992fq,Ikhdair:2003ry}, and the QCD-motivated Coulomb-plus-linear potential (Cor. potential) \cite{Ikhdair:2003ry,Eichten:1978tg,Eichten:1979ms,Eichten:1980mw,Eichten:1995ch}.
The formula and latest values of those wave functions at the origin and their first derivatives at the origin can be found in our earlier work \cite{lx}.
In Tables \ref{wavefunc1}-\ref{wavefunc3}, the contributions from four P-wave states ($^1P_1$, $^3P_J$ with $J=0,1,2$) are summed up.
It is shown that the cross sections change dramatically when we choose different potential models.
For the production of double charmonium $|(c\bar{c})[n]\rangle$~/$|(c\bar{c})[n']\rangle$ in Table \ref{wavefunc1}, we always obtain the minimum under the I.O. potential model, and obtain the maximum under the B.T. potential models.
While for the production of double bottomonium $|(b\bar{b})[n]\rangle$~/$|(b\bar{b})[n']\rangle$ in Table \ref{wavefunc2}, we obtain the minimum under C.K. potential models, and obtain the maximum under the B.T. potential models.
And for the production of double $|(b\bar{c})[n]\rangle$~/$|(c\bar{b})[n']\rangle$ in Table \ref{wavefunc3}, we obtain the minimum under the C.K. potential models, and obtain the maximum under the I.O. potential models.

\begin{table}
\caption{Uncertainties of total cross sections (units:~$\times 10^{-5}~fb$) caused by five different potential models for double charmonium $|(c\bar{c})[n]\rangle$~/$|(c\bar{c})[n']\rangle$ in $e^+e^-\to Z^0\to |(c\bar{c})[n]\rangle+|(c\bar{c})[n']\rangle$ ($n_f=3$).}
\begin{tabular}{|c|c|c|c|c|c|}
\hline
$|(c\bar{c})[n]\rangle$&~B.T.~&~J.~&~I.O.~&~C.K.~&~Cor.~\\
$\sigma_{(|(c\bar{c})[^1S_0]\rangle+|(c\bar{c})[^3S_1]\rangle)}$&~18.95~&~3.927~&~1.001~&~1.653 ~&~2.976~\\
\hline
$\sigma_{(|(c\bar{c})[^3S_1]\rangle+|(c\bar{c})[^3S_1]\rangle)}$&~64.16~&~13.30~&~3.390~&~5.597 ~&~10.07~\\
\hline
$\sigma_{(|(c\bar{c})[^1S_0]\rangle+|(c\bar{c})[^1P_1]\rangle)}$&~60.76~&~14.78~&~2.299~&~4.124~&~6.804~\\
\hline
$\sigma_{(|(c\bar{c})[^3S_1]\rangle+|(c\bar{c})[^1P_1]\rangle)}$&~454.0~&~110.4~&~17.18~&~30.82~&~50.84~\\
\hline
$\sigma_{(|(c\bar{c})[^1S_0]\rangle+|(c\bar{c})[^3P_0]\rangle)}$&~137.0~&~33.32~&~5.183~&~9.299~&~15.34~\\
\hline
$\sigma_{(|(c\bar{c})[^1S_0]\rangle+|(c\bar{c})[^3P_1]\rangle)}$&~18.87~&~4.589~&~0.714~&~1.281~&~2.113~\\
\hline
$\sigma_{(|(c\bar{c})[^1S_0]\rangle+|(c\bar{c})[^3P_2]\rangle)}$&~284.2~&~69.11~&~10.75~&~19.29~&~31.83~\\
\hline
$\sigma_{(|(c\bar{c})[^3S_1]\rangle+|(c\bar{c})[^3P_0]\rangle)}$&~44.03~&~10.71~&~1.666~&~2.989~&~4.931~\\
\hline
$\sigma_{(|(c\bar{c})[^3S_1]\rangle+|(c\bar{c})[^3P_1]\rangle)}$&~17.62~&~4.285~&~0.667~&~1.196~&~1.973~\\
\hline
$\sigma_{(|(c\bar{c})[^3S_1]\rangle+|(c\bar{c})[^3P_2]\rangle)}$&~57.51~&~13.99~&~2.176~&~3.903~&~6.440~\\
\hline
Sum&~1158~&~278.6~&~45.08~&~80.23~&~133.5~\\
\hline\hline
\end{tabular}
\label{wavefunc1}
\end{table}
\begin{table}
\caption{Uncertainties of total cross sections (units:~$\times 10^{-4}~fb$) caused by five different potential models for double bottomonium $|(b\bar{b})[n]\rangle$~/$|(b\bar{b})[n']\rangle$ in $e^+e^-\to Z^0\to |(b\bar{b})[n]\rangle+ |(b\bar{b})[n']\rangle$ ($n_f=4$).}
\begin{tabular}{|c|c|c|c|c|c|}
\hline
$|(b\bar{b})[n]\rangle$&~B.T.~&~J.~&~I.O.~&~C.K.~&~Cor.~\\
\hline\hline
$\sigma_{(|(b\bar{b})[^1S_0]\rangle+|(b\bar{b})[^3S_1]\rangle)}$&~428.6~&~83.47~&~164.3~&~46.30~&~137.8~\\
\hline
$\sigma_{(|(b\bar{b})[^3S_1]\rangle+|(b\bar{b})[^3S_1]\rangle)}$&~119.9~&~23.35~&~45.97~&~12.95~&~38.55~\\
\hline
$\sigma_{(|(b\bar{b})[^1S_0]\rangle+|(b\bar{b})[^1P_1]\rangle)}$&~53.27~&~6.580~&~6.542~&~3.311~&~6.263~\\
\hline
$\sigma_{(|(b\bar{b})[^3S_1]\rangle+|(b\bar{b})[^1P_1]\rangle)}$&~61.46~&~7.591~&~7.547~&~3.820~&~7.226~\\
\hline
$\sigma_{(|(b\bar{b})[^1S_0]\rangle+|(b\bar{b})[^3P_0]\rangle)}$&~9.215~&~1.138~&~1.132~&~0.573~&~1.083~\\
\hline
$\sigma_{(|(b\bar{b})[^1S_0]\rangle+|(b\bar{b})[^3P_1]\rangle)}$&~14.31~&~1.767~&~1.757~&~0.890~&~1.682~\\
\hline
$\sigma_{(|(b\bar{b})[^1S_0]\rangle+|(b\bar{b})[^3P_2]\rangle)}$&~28.15~&~3.477~&~3.457~&~1.750~&~3.310~\\
\hline
$\sigma_{(|(b\bar{b})[^3S_1]\rangle+|(b\bar{b})[^3P_0]\rangle)}$&~68.37~&~8.445~&~8.396~&~4.250~&~8.038~\\
\hline
$\sigma_{(|(b\bar{b})[^3S_1]\rangle+|(b\bar{b})[^3P_1]\rangle)}$&~117.5~&~14.51~&~14.43~&~7.304~&~13.81~\\
\hline
$\sigma_{(|(b\bar{b})[^3S_1]\rangle+|(b\bar{b})[^3P_2]\rangle)}$&~230.6~&~28.48~&~28.32~&~14.33~&~27.11~\\
\hline
Sum&~1132~&~178.9~&~282.1~&~95.55~&~245.1~\\
\hline\hline
\end{tabular}
\label{wavefunc2}
\end{table}
\begin{table}
\caption{Uncertainties of total cross sections (units:~$\times 10^{-3}~fb$) caused by five different potential models for double mesons $|(c\bar{b})[n]\rangle~/|(b\bar{c})[n']\rangle$ in $e^+e^-\to Z^0\to |(c\bar{b})[n]\rangle+ |(b\bar{c})[n']\rangle$ ($n_f=3$).}
\begin{tabular}{|c|c|c|c|c|c|}
\hline
$|(b\bar{b})[n]\rangle$&~B.T.~&~J.~&~I.O.~&~C.K.~&~Cor.~\\
\hline\hline
$\sigma_{(|(c\bar{b})[^1S_0]\rangle+|(b\bar{c})[^3S_1]\rangle)}$&~634.6~&~175.1~&~1653~&~72.88~&~136.2~\\
\hline
$\sigma_{(|(c\bar{b})[^3S_1]\rangle+|(b\bar{c})[^3S_1]\rangle)}$&~1150~&~317.2~&~2996~&~132.1~&~246.9~\\
\hline
$\sigma_{(|(c\bar{b})[^1S_0]\rangle+|(b\bar{c})[^1P_1]\rangle)}$&~4.810~&~2.014~&~8.588~&~0.585~&~0.942~\\
\hline
$\sigma_{(|(c\bar{b})[^3S_1]\rangle+|(b\bar{c})[^1P_1]\rangle)}$&~11.77~&~4.929~&~21.01~&~1.432~&~2.306~\\
\hline
$\sigma_{(|(c\bar{b})[^1S_0]\rangle+|(b\bar{c})[^3P_0]\rangle)}$&~15.27~&~6.394~&~27.26~&~1.858~&~2.991~\\
\hline
$\sigma_{(|(c\bar{b})[^1S_0]\rangle+|(b\bar{c})[^3P_1]\rangle)}$&~6.162~&~2.580~&~11.00~&~0.750~&~1.207~\\
\hline
$\sigma_{(|(c\bar{b})[^1S_0]\rangle+|(b\bar{c})[^3P_2]\rangle)}$&~3.093~&~1.295~&~5.522~&~0.376~&~0.606~\\
\hline
$\sigma_{(|(c\bar{b})[^3S_1]\rangle+|(b\bar{c})[^3P_0]\rangle)}$&~454.2~&~190.2~&~811.0~&~55.27~&~88.98~\\
\hline
$\sigma_{(|(c\bar{b})[^3S_1]\rangle+|(b\bar{c})[^3P_1]\rangle)}$&~270.1~&~113.1~&~482.3~&~32.87~&~52.91~\\
\hline
$\sigma_{(|(c\bar{b})[^3S_1]\rangle+|(b\bar{c})[^3P_2]\rangle)}$&~1133~&~474.4~&~2023~&~137.9~&~222.0~\\
\hline
Sum&~3692~&~1231~&~8062~&~437.1~&~756.9~\\
\hline\hline
\end{tabular}
\label{wavefunc3}
\end{table}

For the uncertainties of total cross sections (units:~$\times 10^{-5}$ fb) caused by the deviation of CM energy $\sqrt{s}$ away from $m_Z$, one can have a visual impression in Figs. \ref{ccds} and \ref{bbds}.
It is shown that the cross sections decreases dramatically with the deviation of CM energy $\sqrt{s}$ away from $m_Z$. To obtain a quantitative impression, we display the uncertainties caused by the deviation of CM energy $\sqrt{s}$ away from $m_Z$ by 1\% and 3\% for the $Z^0$ propagated process in Tables \ref{sqrtsUncertanty1}-\ref{sqrtsUncertanty3}.
\begin{table}
\caption{Uncertainties of total cross sections (units:~$\times 10^{-5}~fb$) caused by the deviation of CM energy $\sqrt{s}$ away from $m_Z$ for double charmonium $|(c\bar{c})[n]\rangle$~/$|(c\bar{c})[n']\rangle$ in $e^+e^-\to Z^0 \to |(c\bar{c})[n]\rangle+|(c\bar{c})[n']\rangle$ under B.T. model ($n_f=3$).}
\begin{tabular}{|c|c|c|c|c|c|}
\hline
~$\sqrt{s}$~&$97\%m_Z$&$99\%m_Z$&$m_Z$&$101\%m_Z$&$103\%m_Z$\\
\hline\hline
$\sigma_{(|(c\bar{c})[^1S_0]\rangle+|(c\bar{c})[^3S_1]\rangle)}$&~3.778~&~12.87~&18.95&~11.83~&~2.830~\\
\hline
$\sigma_{(|(c\bar{c})[^3S_1]\rangle+|(c\bar{c})[^3S_1]\rangle)}$&~12.79~&~43.57~&64.16&~40.06~&~9.582~\\
\hline
$\sigma_{(|(c\bar{c})[^1S_0]\rangle+|(c\bar{c})[^1P_1]\rangle)}$&~11.39~&~40.43~&60.76&~38.70~&~9.626~\\
\hline
$\sigma_{(|(c\bar{c})[^3S_1]\rangle+|(c\bar{c})[^1P_1]\rangle)}$&~85.60~&~302.6~&454.0&~288.6~&~71.55~\\
\hline
$\sigma_{(|(c\bar{c})[^1S_0]\rangle+|(c\bar{c})[^3P_0]\rangle)}$&~25.67~&~91.12~&137.0&~87.23~&~21.70~\\
\hline
$\sigma_{(|(c\bar{c})[^1S_0]\rangle+|(c\bar{c})[^3P_1]\rangle)}$&~3.760~&~12.81~&18.87&~11.78~&~2.818~\\
\hline
$\sigma_{(|(c\bar{c})[^1S_0]\rangle+|(c\bar{c})[^3P_2]\rangle)}$&~53.39~&~189.2~&284.2&~180.9~&~44.94~\\
\hline
$\sigma_{(|(c\bar{c})[^3S_1]\rangle+|(c\bar{c})[^3P_0]\rangle)}$&~8.362~&~29.42~&44.03&~27.92~&~6.892~\\
\hline
$\sigma_{(|(c\bar{c})[^3S_1]\rangle+|(c\bar{c})[^3P_1]\rangle)}$&~3.512~&~11.96~&17.62&~11.00~&~2.630~\\
\hline
$\sigma_{(|(c\bar{c})[^3S_1]\rangle+|(c\bar{c})[^3P_2]\rangle)}$&~11.16~&~38.70~&57.51&~36.22~&~8.821~\\
\hline
Sum&~219.4~&~772.7~&~1157~&~734.2~&~181.4~\\
\hline\hline
\end{tabular}
\label{sqrtsUncertanty1}
\end{table}
\begin{table}
\caption{Uncertainties of total cross sections (units:~$\times 10^{-4}~fb$) caused by the deviation of CM energy $\sqrt{s}$ away from $m_Z$ for double bottomonium $|(b\bar{b})[n]\rangle$~/$|(b\bar{b})[n']\rangle$  in $e^+e^-\to Z^0 \to |(b\bar{b})[n]\rangle+|(b\bar{b})[n']\rangle$ under B.T. model ($n_f=4$).}
\begin{tabular}{|c|c|c|c|c|c|}
\hline
~$\sqrt{s}$~&$97\%m_Z$&$99\%m_Z$&$m_Z$&$101\%m_Z$&$103\%m_Z$\\
\hline\hline
$\sigma_{(|(b\bar{b})[^1S_0]\rangle+|(b\bar{b})[^3S_1]\rangle)}$&85.10&290.7&428.6&~267.9~&~64.21~\\
\hline
$\sigma_{(|(b\bar{b})[^3S_1]\rangle+|(b\bar{b})[^3S_1]\rangle)}$&23.73&81.22&119.9&~74.98~&~18.00~\\
\hline
$\sigma_{(|(b\bar{b})[^1S_0]\rangle+|(b\bar{b})[^1P_1]\rangle)}$&9.950&35.40&53.27&~33.96~&~8.467~\\
\hline
$\sigma_{(|(b\bar{b})[^3S_1]\rangle+|(b\bar{b})[^1P_1]\rangle)}$&11.82&41.24&61.46&~38.82~&~9.501~\\
\hline
$\sigma_{(|(b\bar{b})[^1S_0]\rangle+|(b\bar{b})[^3P_0]\rangle)}$&1.708&6.109&9.215&~5.890~&~1.475~\\
\hline
$\sigma_{(|(b\bar{b})[^1S_0]\rangle+|(b\bar{b})[^3P_1]\rangle)}$&2.834&9.695&14.31&~8.948~&~2.148~\\
\hline
$\sigma_{(|(b\bar{b})[^1S_0]\rangle+|(b\bar{b})[^3P_2]\rangle)}$&5.340&18.80&28.15&~17.86~&~4.411~\\
\hline
$\sigma_{(|(b\bar{b})[^3S_1]\rangle+|(b\bar{b})[^3P_0]\rangle)}$&13.36&46.12&68.37&~42.96~&~10.41~\\
\hline
$\sigma_{(|(b\bar{b})[^3S_1]\rangle+|(b\bar{b})[^3P_1]\rangle)}$&23.37&79.71&117.5&~73.37~&~17.57~\\
\hline
$\sigma_{(|(b\bar{b})[^3S_1]\rangle+|(b\bar{b})[^3P_2]\rangle)}$&45.46&156.2&230.6&~144.5~&~34.82~\\
\hline
Sum&~222.7~&~765.2~&~1131~&~709.2~&~171.0~\\
\hline\hline
\end{tabular}
\label{sqrtsUncertanty2}
\end{table}
\begin{table}
\caption{Uncertainties of total cross sections (units:~$\times 10^{-3}~fb$) caused by the deviation of CM energy $\sqrt{s}$ away from $m_Z$ for double mesons $|(c\bar{b})[n]\rangle~/|(b\bar{c})[n']\rangle$ in $e^+e^-\to Z^0 \to |(c\bar{b})[n]\rangle+|(b\bar{c})[n']\rangle$ under B.T. model  ($n_f=3$).}
\begin{tabular}{|c|c|c|c|c|c|}
\hline
~$\sqrt{s}$~&$97\%m_Z$&$99\%m_Z$&$m_Z$&$101\%m_Z$&$103\%m_Z$\\
\hline
$\sigma_{(|(c\bar{b})[^1S_0]\rangle+|(b\bar{c})[^3S_1]\rangle)}$&~121.9~&~425.6~&634.6&~401.0~&~98.27~\\
\hline
$\sigma_{(|(c\bar{b})[^3S_1]\rangle+|(b\bar{c})[^3S_1]\rangle)}$&~220.3~&~770.8~&1150&~727.5~&~178.6~\\
\hline
$\sigma_{(|(c\bar{b})[^1S_0]\rangle+|(b\bar{c})[^1P_1]\rangle)}$&~0.910~&~3.210~&4.810&~3.054~&~0.755~\\
\hline
$\sigma_{(|(c\bar{b})[^3S_1]\rangle+|(b\bar{c})[^1P_1]\rangle)}$&~2.322~&~7.964~&11.77&~7.371~&~1.774~\\
\hline
$\sigma_{(|(c\bar{b})[^1S_0]\rangle+|(b\bar{c})[^3P_0]\rangle)}$&~2.851~&~10.15~&15.27&~9.739~&~2.429~\\
\hline
$\sigma_{(|(c\bar{b})[^1S_0]\rangle+|(b\bar{c})[^3P_1]\rangle)}$&~1.226~&~4.181~&6.162&~3.849~&~0.922~\\
\hline
$\sigma_{(|(c\bar{b})[^1S_0]\rangle+|(b\bar{c})[^3P_2]\rangle)}$&~0.611~&~2.094~&3.093&~1.937~&~0.466~\\
\hline
$\sigma_{(|(c\bar{b})[^3S_1]\rangle+|(b\bar{c})[^3P_0]\rangle)}$&~81.63~&~298.0~&454.2&~293.3~&~75.03~\\
\hline
$\sigma_{(|(c\bar{b})[^3S_1]\rangle+|(b\bar{c})[^3P_1]\rangle)}$&~51.16~&~180.3~&270.1&~171.5~&~42.40~\\
\hline
$\sigma_{(|(c\bar{b})[^3S_1]\rangle+|(b\bar{c})[^3P_2]\rangle)}$&~20.57~&~745.6~&1133&~729.0~&~185.3 ~\\
\hline
Sum&~503.5~&~2448~&~3683~&~2348~&~585.9~\\
\hline\hline
\end{tabular}
\label{sqrtsUncertanty3}
\end{table}

\section{Conclusions}
In the present work, we make a comprehensive study on the production of double excited quarkonium $|(c\bar{c})[n]\rangle$~/$|(c\bar{c})[n']\rangle$, $|(b\bar{b})[n]\rangle$~/$|(b\bar{b})[n']\rangle$, and $|(b\bar{c})[n]\rangle~/|(c\bar{b})[n']\rangle$ through $e^+e^-\to \gamma^*/Z^0\to |(Q\bar{Q'})[n]\rangle+ |(Q'\bar{Q})[n']\rangle$ under the NRQCD factorization framework at the future $Z$ factory, where the $[n]$~/$[n']$ represents the color-singlet $[^1S_0]\rangle, ~[^3S_1]\rangle, ~[^1P_1]\rangle$, and $[^3P_J]\rangle$ ($J=0,1,2$) heavy quarkonium.
The Dirac matrices at the amplitude level is disposed with the``improved trace technology", which is helpful for deriving compact analytical results especially for the complicated $P$-wave processes with massive spinors.
The total cross sections $\sigma(\sqrt{s})$ and differential distributions $d\sigma/dcos\theta$ and $d\sigma/dp_{t}$ for all the double excited heavy quarkonium ($n=1$) are studied in detail.
For a sound estimation, we further study the uncertainties of the cross sections caused by the varying mass of $c$- and $b$-quarks, the non-perturbative matrix elements under five potential models, and deviation of CM energy $\sqrt{s}$ away from $m_Z$ by 1\% and 3\%.

In addition to the ground states, it is found that the production rates of double excited charmonium, double excited bottomonium, and double excited $Bc$ mesons are considerable in the processes of $e^+e^-\to \gamma^*/Z^0\to |(Q\bar{Q'})[n]\rangle+ |(Q'\bar{Q})[n']\rangle$ at the super $Z$ factory with high luminosity ${\cal L}\approx 10^{36}cm^{-2}s^{-1}$.
Then, such the super $Z$ factory could provide a useful platform to study on the double excited charmonium,  the double excited bottomonium, and the double excited $Bc$ mesons.
In addition, we find that cross sections change dramatically when adopting different potential models, which would be the major source of uncertainty.
And the deviation of CM energy $\sqrt{s}$ away from $Z^0$ pole at the future super $Z$ factory will also have great influence on the production rates.

\hspace{2cm}

{\bf Acknowledgements}:
This work was supported in part by
the National Natural Science Foundation of China under Grant No. 11905112,
and the Natural Science Foundation of Shandong Province under Grant No. ZR2019QA012.

\end{document}